\documentclass[aps,pre,reprint,twocolumn,superscriptaddress,showpacs]{revtex4-1}
\usepackage{amssymb,amsmath}
\usepackage[table]{xcolor}
\usepackage{graphicx}
\usepackage{mathtools}
\usepackage[export]{adjustbox}
\usepackage{overpic}
\usepackage[colorlinks=true,
 linkcolor=black,
 urlcolor=blue,
 citecolor=blue]{hyperref}
\usepackage{nicefrac}
\usepackage{multirow}
\usepackage{lipsum} 
\usepackage[normalem]{ulem}
\usepackage{pbox}
\newcolumntype{C}[1]{>{\centering\let\newline\\\arraybackslash\hspace{0pt}}m{#1}}

\usepackage{enumitem}

\usepackage{framed,color}
\definecolor{shadecolor}{rgb}{0.85,0.80,0.80}

\definecolor{myorange}{RGB}{253, 184, 99}
\definecolor{mypurple}{RGB}{178, 171, 210}

\newcommand{\comments}[1]{}
\usepackage{multirow}

\definecolor{ao}{rgb}{0.0, 0.4, 0.1}

\usepackage[normalem]{ulem}

\newcommand{\removed}[1]{}

\newcommand{\be}{\begin{equation}}
\newcommand{\ee}{\end{equation}}
\newcommand{\bd}{\begin{displaymath}}
\newcommand{\ed}{\end{displaymath}}
\newcommand{\BE}{\begin{eqnarray}}
\newcommand{\EE}{\end{eqnarray}}

\newcommand{\boldrho}{{\text{\boldmath $\rho$}}}

\newcommand{\avg}[1]{\left\langle{#1}\right\rangle}

\begin{document}

\title{Effective diffusion coefficients in reaction-diffusion systems with anomalous transport}
\author{Joseph W. Baron}
\email{joseph.baron@postgrad.manchester.ac.uk}
\affiliation{Theoretical Physics, School of Physics and Astronomy,
The University of Manchester, Manchester M13 9PL, United Kingdom}

\author{Tobias Galla}
\email{tobias.galla@manchester.ac.uk}
\affiliation{Theoretical Physics, School of Physics and Astronomy,
The University of Manchester, Manchester M13 9PL, United Kingdom}
 
\begin{abstract}
We show that the Turing patterns in reaction systems with subdiffusion can be replicated in an effective system with Markovian cross-diffusion. The effective system has the same Turing instability as the original system, and the same patterns. If particles are short-lived, the transient dynamics are captured as well. We use the cross-diffusive system to define effective diffusion coefficients for the system with anomalous transport, and we show how they can be used to efficiently describe the Turing instability. We also demonstrate that the mean squared displacement of a suitably defined ensemble of subdiffusing particles grows linearly with time, with a diffusion coefficient which agrees with our earlier calculations. We verify these deductions by numerically integrating both the fractional reaction-diffusion equation and its normally diffusing counterpart. Our findings suggest that cross-diffusive behaviour can come about as a result of anomalous transport.

\end{abstract}

\maketitle


 \section{Introduction}
In his seminal paper \cite{turingchemical} Turing provided a simple mechanism for spontaneous pattern formation in chemical reaction systems. Since this work, the reaction-diffusion paradigm \cite{crosshohenberg} has been used ubiquitously as a tool for the description of pattern-forming phenomena, ranging from predator-prey interactions \cite{brittonbook} to developmental biology \cite{kondo,murrayII}.

In this context, one typically assumes that constituents diffuse in a manner such that the mean squared displacement of individuals grows linearly with time. Mathematically, $\langle {\bf r}^2 \rangle \sim t$, where ${\bf r}$ denotes the position vector of a typical particle, and where the angular brackets define an average over trajectories. The rate of diffusion can then be quantified by means of the diffusion constant $D = \langle {\bf r}^2 \rangle/(2t)$ \cite{smoluchowski,einstein}. In one spatial dimension this simplifies to $D = \langle x^2 \rangle/(2t)$. However, the situation may not always be this simple. Due to the many complicated interactions between the reactants and their environment, the diffusion may be of an `anomalous' nature. For example, the particles or individuals involved may be subject to `trapping' effects, which can result in subdiffusive behaviour: $\langle x^2 \rangle \sim t^\alpha$ where $0<\alpha < 1$ \cite{randomwalkguide, mendezfedotov, anomaloustransport}. Reaction-diffusion equations with anomalous (or `non-Fickian') diffusion have been used to describe systems ranging from the neolithic transition in Europe \cite{vladross,neolithic} to photoluminescence in semi-conductors \cite{subdiffsemicond} to morphogen gradient formation \cite{yustabadlindenberg, fedotovfalconer2}.

How, then, might one go about defining a quantity which describes the speed of anomalous diffusion? What is the analogue of the diffusion constant $D$? In order to be useful, such a quantity must not only inform us as to how quickly different reactant types diffuse, but it must also be of physical relevance and have predictive power. For example, a criterion for pattern formation in normally diffusive reaction-diffusion systems can be written in terms of the diffusion constants and the other system parameters \cite{crossbook}. So, by introducing an `effective diffusion coefficient', we might aim to produce a similar criterion for pattern formation in systems with anomalous diffusion.

The notion of an effective diffusion coefficient has been utilised previously in the study of transport in confined geometries \cite{confinedgeometries} and for tilted periodic potentials \cite{reimannporous,reimannporous2}. The possibility of defining an effective diffusion constant in a reaction-subdiffusion system was mentioned in \cite{yadavmilu}. In this paper we present a more detailed argument for the definition of effective diffusion coefficients, and seek to interpret the results.

We approach the problem as follows. We start from a reaction-diffusion system in which the reactants can exhibit subdiffusive behaviour. We refer to this as the `original' system. We then derive an `effective' normally diffusing reaction-diffusion system, which exhibits pattern formation for the same sets of model parameters as the original anomalously diffusing system. Furthermore, we argue that the normally diffusing system gives rise to the same patterns as the original system in the long-term and, under certain conditions, replicates the transient dynamics of the original system to a good approximation. It is the diffusion coefficients in this normally diffusing system which we use to characterise the pattern formation in the original subdiffusive system. Although we use subdiffusion as the primary example in this work, our method could be applied to other varieties of non-Fickian diffusion than just the subdiffusive kind. 

We find that the effective diffusion in reaction-diffusion systems with anomalous transport is in general cross-diffusive. That is, the diffusion of one substance may be affected by a concentration gradient in another. As such, we suggest that anomalous transport can be a possible mechanism for the production of cross-diffusive effects in experiments, in addition to those that are already known. 

We also show that the collective behaviour of constituents in the stationary state can give rise to a mean squared displacement which scales linearly with time, $\langle x^2 \rangle \sim t$, despite the underlying subdiffusive motion of the individuals. To demonstrate this, we consider an ensemble in which particles are subject to diffusion and removal events but are replaced upon removal, thus keeping the ensemble size the same, as would be the case in the homogeneous steady state. The effective diffusion coefficient found from this reasoning agrees with our earlier definitions. 

The remainder of this paper is set out as follows: In Section \ref{section:model}, we present the necessary background information. In particular, we summarise the main features of reaction-subdiffusion systems, which we use in the rest of the work. In Section \ref{section:effectivesystem}, we construct the corresponding `effective' normally diffusing system. We go on to show that the condition for Turing pattern formation in the subdiffusive system can be written in terms of the effective diffusion coefficients and that the normally diffusive system produces the same patterns as the original system. Additionally, we show that, under certain circumstances, the normally diffusing system approximates the time-dependent behaviour of the original system. The interpretation of the effective diffusion coefficients in terms of the underlying microscopic processes is also discussed. In Section \ref{section:simulations}, we confirm the theoretical results by numerical integration of the reaction-subdiffusion equation, and of its normally diffusing counterpart. Finally, we summarise and discuss our findings in Section \ref{section:discussion}. 


\section{Model definition and background}\label{section:model}
In this section, we discuss the formulation of reaction-diffusion systems in terms of a continuous-time random walk (CTRW) model. The CTRW approach is a standard method by which to derive reaction-diffusion equations with anomalous diffusive behaviour \cite{mendezfedotov,randomwalkguide,anomaloustransport}. We also discuss the stability of the homogeneous steady states in such systems.
\subsection{General reaction-diffusion model}
We consider a system in one spatial dimension which involves different types of particles. These particles may hop from location to location and be involved in reactions. When a reaction occurs, particles of each flavour can be created or destroyed. In principle this defines a stochastic system if the overall particle number is finite. In this paper, however, we ignore the stochastic fluctuations and examine only the deterministic behaviour. In doing so, we are assuming that the total number of particles is sufficiently large, so that the fluctuations are comparatively negligible. A further convenient consequence of dealing with large particle numbers is that the discrete quantity of particles can be approximated as a continuous concentration. 

Within this approximation, the local particle number density (or concentration) of a particular species $i$ at a time $t$ and at location $x$ is denoted by $\rho_i\left(x, t\right)$. This means that $\int_{x_1}^{x_2} \rho_i\left(x, t\right)dx$ is the number of particles of type $i$ in the region $x_1<x<x_2$. We write $\boldrho$ for the vector $(\rho_1, \rho_2,\dots)$, i.e., $\boldrho(x,t)$ indicates the concentrations of all particles types at a given location and time. Given the initial location of a particle of type $i$, a waiting time is drawn from a distribution $\psi_i\left(t\right)$. Waiting times for each particle are drawn independently. We do not specify these distributions at this point for the purposes of generality. Each particle then remains where it is until it has waited the allotted time, at which point it hops by a distance $x$, drawn from a hopping kernel $\phi_i\left(x\right)$. The spatial coordinate $x$ can be continuous or can denote a position on a discrete lattice. Once the particle has hopped, a new waiting time is drawn and the process repeats itself. 

Given these definitions, one can define the hazard rate of hopping $h_i\left(t\right) = \psi_i\left(t\right)/\Psi_i\left(t\right)$, where $\Psi_i\left(t\right) = \int_t^\infty \psi_i\left(\tau\right) d\tau$ is the probability that a particle has not hopped for a time $t$ since its last hop. The hazard rate characterises the proclivity of particles to hop a time $t$ after their last hop. It is constant in the case of regular diffusion but time-dependent in the general case.

Particles may also be involved in reactions during their sojourn periods at a fixed location. We label these different types of reactions with an index $r$. They occur with rates which depend on the concentration of particles at that location. The result of a reaction is the local production and/or annihilation of particles. We denote the rate at which a reaction of type $r$ occurs by $\lambda_r\left[\boldrho\left(x,t\right)\right]$ and the number of particles of type $i$ which are produced or annihilated in such a reaction by $\nu_i^r$. If particles are annihilated, scheduled hopping events involving these particles no longer occur. Particles which are involved in reactions have their waiting-times redrawn; this approach applies to irreversible reactions, but special consideration has to be given in the case where the reactions are reversible \cite{sokolov2}. We do not consider such reaction schemes in this paper.

In order to deal with the non-constant hazard rates $h_i\left(t\right)$ and to derive an equation which describes the time evolution of the concentrations $\boldrho(x,t)$, one can introduce an additional temporal coordinate $\tau$ -- the time since the last hop for an individual particle. The concentrations can then be subdivided by introducing the quantities $\boldrho(x,\tau,t)$ where $\boldrho(x,t) = \int_0^t \boldrho(x,\tau,t) d\tau$. Using such a coordinate, the problem is recast as Markovian. One is then able to derive a generalised reaction-diffusion equation. A detailed derivation of equations of this type can be found in \cite{yadav,vladross}. Carrying out a calculation along these lines, one obtains
\begin{widetext}
\begin{align}
\frac{\partial \rho_i\left(x,t\right)}{\partial t} &= \int dx'~\left[\phi_i\left(x-x'\right) - \delta\left(x-x'\right) \right] \int_0^t d\tau~ K_i\left(\tau\right)\rho_i\left(x',  t-\tau \right) e^{ -\int_{t-\tau}^t dt'~  \frac{R^-_i\left[\boldrho\left(x',t' \right)\right]}{\rho_i\left(x',t' \right)} } \nonumber \\
&+ R^+_i\left[\boldrho\left(x,t\right)\right] -R^-_i\left[\boldrho\left(x,t\right)\right] .\label{generalreactdiff}
\end{align}
\end{widetext}
We have defined the following objects:
\BE
R_i^+\left[\boldrho\left(x,t\right)\right] &= &\sum_r H\left(\nu_i^r\right) \lambda_r\left[\boldrho\left(x,t\right)\right] \nu^r_i, \nonumber \\
R_i^-\left[\boldrho\left(x,t\right)\right] &=& \sum_r H\left(-\nu_i^r\right) \lambda_r\left[\boldrho\left(x,t\right)\right] \nu^r_i, 
 \label{eq:def}
\EE
where $H(u)=1$ if $u\geq 0$ and $H(u)=0$ otherwise.
The quantities $R_i^\pm$ are the net production and annihilation rates, respectively, of particles of type $i$. In Eq. (\ref{generalreactdiff}) $K_i\left(\tau\right)$ is the `memory kernel' defined through its Laplace transform,
\begin{align}
\mathcal{L}_t\left\{ K_i\left(t\right)\right\} = \frac{\mathcal{L}_t\left\{ \psi\left(t\right)\right\}}{\mathcal{L}_t\left\{ \Psi\left(t\right)\right\}}. \label{memorykernel}
\end{align} 
The Laplace transform is given by $\mathcal{L}_t\left\{ f\left(t\right)\right\}\left(u\right)=\int_0^\infty e^{-ut }f\left(t\right)dt $. \\
In Eq.~(\ref{generalreactdiff}), the reaction term, $R^+\left(\boldrho\right) - R^-\left(\boldrho\right)$, and the diffusion term [which contains convolutions with both the hopping kernel $\phi\left(x\right)$ and the memory kernel $K_i\left(t\right)$] are coupled via the exponential factor in the convolution with the memory kernel. This coupling arises from the fact that particles may be annihilated before they hop: If the death rate $R^-_i$ were identically zero, the contribution of the concentration at a time $t-\tau$ to the integral over time in Eq.~(\ref{generalreactdiff}) would be $K_i\left(\tau\right)\rho_i\left(x',  t-\tau \right)$. To account for a non-zero death rate, this contribution is weighted by the probability that a particle of type $i$ survives the time interval from $t-\tau$ to $t$; this probability is given by $ e^{ -\int_{t-\tau}^t dt'~  \frac{R^-_i\left[\boldrho\left(x',t' \right)\right]}{\rho_i\left(x',t' \right)} }$. This coupling term is absent in the case of Markovian hopping. That is, if we choose the exponential waiting time distribution $\psi_i\left(t\right) = \frac{1}{\tau_i} e^{-\frac{t}{\tau_i}}$ we obtain $K_i\left(t\right) = \frac{1}{\tau_i}\delta\left(t\right)$, and the reaction and diffusion terms in Eq.~(\ref{generalreactdiff}) decouple. In general though, the time-evolution of the concentrations depends on the history of the system, and the process is manifestly non-Markovian. 
\subsection{Reaction-subdiffusion system}
In this paper, we will primarily concern ourselves with a specific case of non-Markovian transport: subdiffusion. We choose the waiting-time distribution to be the Mittag-Leffler distribution \cite{gorenflo}
\begin{align}
\psi_i\left(t\right) = -\frac{d}{dt}E_{\alpha_i} \big[- \left(t/\tau_i \right)^{\alpha_i}\big],\label{mittagleffler}
\end{align}
where $0 < \alpha_i <1$ and where $E_{\alpha_i}\left(\cdot\right)$ is the one-parameter Mittag-Leffler function \cite{gorenflobook}. This function can be thought of as a generalised exponential function and is defined as
\begin{align}
E_{\alpha_i}\left(x\right) = \sum_{k=0}^\infty \frac{x^k}{\Gamma\left(\alpha_i k + 1\right)},
\end{align}
where $\Gamma\left(\cdot\right)$ denotes the standard gamma function. The constant $\tau_i$ has units of time; its role is to fix the overall time scale of the hopping process. It is important to note that the Mittag-Leffler distribution has the `heavy-tail' property $\lim_{T \to \infty}\int_0^T t \psi_i\left(t\right)dt \to \infty$. That is, in absence of annihilation the first moment of the waiting times between hops diverges. This makes the Mittag-Leffler function suitable for modelling the hopping of particles subject to `trapping' effects \cite{randomwalkguide}. 

We allow the hopping kernel to take a general symmetrical form. If one presumes that the hopping kernel is of the form $\phi_i\left(x\right) = \Phi_i\left(x/\xi\right)$, where $\xi$ characterises the typical hopping distance and $\Phi_i\left(\cdot\right)$ is even, the Fourier transform of the hopping kernel has following asymptotics 
\begin{align}
\tilde \phi_i\left(k\right) \approx 1 - \sigma_i^2 \xi^2 k^2 +{\cal O}(\xi^4), \label{hoppingkernel}
\end{align}
for small $\xi$, where $\sigma_i$ is a constant related to the variance of the hopping distances such that $-\frac{\partial^2}{\partial k^2} \tilde\phi\left(k\right) \vert_{k = 0} = 2 \sigma_i^2 \xi^2$.

If one were to track the movement of an ensemble of particles which hopped with waiting-times drawn from the distribution in Eq.~(\ref{mittagleffler}) and hopping distances drawn from a hopping kernel described by Eq.~(\ref{hoppingkernel}), one would obtain subdiffusive behaviour, characterised by the mean squared displacement $\langle x^2\rangle \sim t^{\alpha_i}$ \cite{randomwalkguide}. This can be shown using the Montroll-Weiss formula \cite{montrollweiss}. The parameter $\alpha_i$ characterises the degree to which the diffusion is subdiffusive, with $\alpha_i \to 0$ describing extremely subdiffusive behaviour and $\alpha_i = 1$ leading to normal diffusion. We refer to $\alpha_i$ as the anomalous exponent.

The Laplace transform of the memory kernel [see Eq.(\ref{memorykernel})] for the distribution in Eq.~(\ref{mittagleffler}) has the following simple form \cite{fedotovfalconer1},
\begin{align}
\mathcal{L}_t\left\{ K_i\left(t\right)\right\}\left(u\right) = \frac{u^{1-\alpha_i}}{\tau_i^{\alpha_i}} . \label{memorykernelML}
\end{align}
However, the Laplace transform for this function cannot easily be inverted. This makes it difficult to find the memory kernel $K_i(t)$ in the time domain in explicit form. As a result, it is not convenient to write the reaction-diffusion equation in the form of Eq.~(\ref{generalreactdiff}). Instead, we use the Riemann-Liouville fractional derivative, defined as \cite{podlubnybook}
\begin{equation}
{}_0D_t^{1-\alpha} f\left(t\right)= \frac{1}{\Gamma\left(\alpha\right)} \frac{\partial}{\partial t} \int_0^t \frac{f\left(t'\right)}{\left(t-t'\right)^{1-\alpha}}dt' .\label{fracdiffdef}
\end{equation}
This has the following property
\begin{equation}
\mathcal{L}_t\left\{ {}_0D_t^{1-\alpha} f\left(t\right)\right\}\left(u\right)= u^{1-\alpha} \mathcal{L}_t \{ f \}(u), 
\end{equation}
prodivded that $\lim_{t \to 0} \int_0^t f\left(\tau\right) \left(t-\tau\right)^{\alpha -1} d\tau = 0$, which is satisfied if $f\left(t\right)$ is continuous and $f'\left(t\right)$ is integrable \cite{podlubnybook}. It is reasonable to assume that this condition applies to the functions on which the fractional derivative acts in this work, as these functions are related to particle concentrations.

Using these definitions, we define the re-scaled time $\eta_i =  \tau_i\xi^{-2/\alpha_i}$. This re-scaling hints at the underlying subdiffusive nature of the transport. Taking the limit $\xi\to 0$, Eq. (\ref{generalreactdiff}) can be written as
\begin{align}
\frac{\partial \rho_i\left(x,t\right)}{\partial t} &= \frac{\sigma^2_i}{\eta_i^{\alpha_i}}\frac{\partial^2}{\partial x^2} M_i\left(x,t\right) + f_i\left(x,t\right), \label{full}
\end{align}
where $f_i \equiv R^+_i\left[\boldrho\left(x,t\right)\right] -R^-_i\left[\boldrho\left(x,t\right)\right]$ is the total reaction rate, and where we define
\begin{align}
M_i&\left(x,t\right) = 
e^{- \int_0^t \frac{R^-_i \left[\boldrho\left(x,t' \right)\right]}{\rho_i\left(x,t' \right)} dt'} \nonumber \\
&~~~\times {}_0 D_t^{1-\alpha_i} \left\{e^{ \int_0^t \frac{R^-_i \left[\boldrho\left(x,t' \right)\right]}{\rho_i\left(x,t' \right)} dt'} \rho_i\left(x,t \right)\right\} .
\end{align}
The history-dependent quantity $M_i\left(x,t\right)$ characterises the current number of particles of type $i$ hopping away from location $x$ per unit time. References \cite{yustabadlindenberg, vladross, yadav} contain a more detailed discussion of the origins of Eq.~(\ref{full}). 
In a similar way to Eq. (\ref{generalreactdiff}), the reaction and diffusion terms in Eq.~(\ref{full}) are coupled through exponential terms in the fractional derivative. We note that the evolution of the concentrations described by Eq.~(\ref{full}) is dependent on the history of the system and is therefore non-Markovian. However, for $\alpha_i = 1$ the Mittag-Leffler distribution Eq.~(\ref{mittagleffler}) reduces to an exponential distribution, and one obtains the normal reaction-diffusion equation
\begin{align}
\frac{\partial \rho_i\left(x,t\right)}{\partial t} &= \frac{\sigma_i^2}{\eta_i} \frac{\partial^2}{\partial x^2} \rho_i\left(x,t\right) + f_i\left(x,t\right). \label{normal}
\end{align}
In this case the length of time since the last hop is irrelevant to the current hopping rate, and the process is Markovian. \\
We note that Eq.~(\ref{full}) can also be used as an approximation to the dynamics when a general class of heavy-tailed waiting-time distributions is used, not just the Mittag-Leffler function \cite{randomwalkguide, yustabadlindenberg,mendezfedotov}. 


\subsection{Linear stability and pattern formation in the reaction-subdiffusion system}\label{section:stabilityanalysis}
Now that we have discussed the general formulation of reaction-subdiffusion systems, we consider the stability of their fixed points, and discuss conditions for pattern formation. From here on in, we will focus on systems which have two species of reactants.

A stable fixed point in a non-spatial system can become unstable with the introduction of diffusive proceses. As a result of the instability, a limited band of Fourier modes with non-zero wavenumbers can become excited, resulting in the formation of stationary patterns \cite{crosshohenberg,crossbook}. Turing \cite{turingchemical} demonstrated this effect in reaction-diffusion systems with normal diffusion, such as the one described by Eq.~(\ref{normal}), and established a condition for such an instability to occur.

A similar instability has been found in the reaction-subdiffusion system whose evolution is described by Eq.~(\ref{full}), see \cite{yadav,yadavmilu}. We denote the homogeneous fixed point of this system by $\boldrho^0$, which is defined by $f_i(\boldrho^0) = 0$ for all $i$. Writing the derivatives of $f_i$ with respect to particle concentrations as $f_{ij} = \frac{\partial f_i\left(\boldrho^0\right)}{\partial \rho^0_j}$, the stability of this fixed point against uniform perturbations requires \cite{crosshohenberg,crossbook}
\begin{align}
f_{11} + f_{22} &< 0 , \nonumber \\
f_{11}f_{22} - f_{12}f_{21} &>0. \label{homogeneousstability}
\end{align}
Unless indicated otherwise the $f_{ij}$ are always to be evaluated at the homogeneous fixed point.

 We denote deviations from the homogeneous fixed point by $\delta \rho_i\left(x,t\right) \equiv \rho_i\left(x,t\right) - \rho^0_i$. Writing $\delta \tilde{\rho}_i\left(k,t\right)$ for the Fourier transform with respect to the spatial coordinate, one finds the following dynamics for the Fourier mode with wavenumber $k$,\begin{widetext}
\begin{align}
\frac{\partial \delta\tilde\rho_i\left(k,t\right)}{\partial t} = -k^2\frac{\sigma^2_i}{\eta_i^{\alpha_i}}e^{-p_i t} \Bigg[ &  \sum_jA_{ij} \rho_i^0 ~{}_0D_t^{1-\alpha_i} \left\{ e^{p_i t} \int_0^t \delta\tilde\rho_j\left(k,t'\right) dt'\right\} -\sum_j A_{ij} \int_0^t \delta \tilde\rho_j\left(k,t'\right) dt' ~{}_0D_t^{1-\alpha_i}\left\{ e^{p_i t} \rho_i^0\right\} \nonumber \\
&+ {}_0D_t^{1-\alpha_i} \left\{ e^{p_i t} \delta \tilde\rho_i\left(k,t\right)\right\}\Bigg] + \sum_jf_{ij}\delta\tilde\rho_j\left(k,t\right) ,\label{linearisedfrac}
\end{align}
\end{widetext}
to linear order in the perturbations. We have written $p_i= R^-_i\left(\boldrho^0\right)/\rho^0_i$ for the per capita removal rate of species $i$ at the fixed point.  We have also introduced $A_{ij}= \frac{\partial p_i\left(\boldrho^0\right)}{\partial \rho^0_j}$. To keep the notation compact we have omitted the argument $\boldrho^0$ in Eq. (\ref{linearisedfrac}); it is to be understood that the $A_{ij}$ are evaluated at the homogeneous fixed point.  

In \cite{yadav} a condition for the Fourier mode $k$ to be unstable to perturbations was derived for a system with one subdiffusing species and one normally diffusing species. A generalised version of this result for the case where both species subdiffuse is  
\begin{widetext}
\begin{align}
&\left[ f_{11}  -k^2 \frac{\sigma^2_1}{\eta_1^{\alpha_1}}\left( p_1^{1-\alpha_1}+ A_{11} \rho_1^0 p_1^{-\alpha_1} \left(1-\alpha_1\right)\right) \right]\left[f_{22}  -k^2 \frac{\sigma^2_2}{\eta_2^{\alpha_2}}\left( p_2^{1-\alpha_2}+ A_{22} \rho_2^0 p_2^{-\alpha_2} \left(1-\alpha_2\right)\right) \right] \nonumber \\
&-\left[ f_{12}  -k^2 \frac{\sigma^2_1}{\eta_1^{\alpha_1}}\left(  A_{12} \rho_1^0 p_1^{-\alpha_1} \left(1-\alpha_1\right)\right) \right]\left[ f_{21}  -k^2 \frac{\sigma^2_2}{\eta_2^{\alpha_2}}\left(  A_{21} \rho_2^0 p_2^{-\alpha_2} \left(1-\alpha_2\right)\right) \right] < 0 . \label{stabilitycondition}
\end{align}
\end{widetext}
A derivation of this formula is detailed in Appendix \ref{appendix:stabilitycriterion}; we use an alternative route to that of \cite{yadav}. We note that a sign error was made in the calculation in \cite{yadav}. This mistake has been corrected in Eq. (\ref{stabilitycondition}), see also Appendix \ref{yadavcorrection}.



\section{Construction of the effective normally diffusive system}\label{section:effectivesystem}
\subsection{Definition of the system and stationary patterns}\label{section:deffeff}

In the following, we construct a Markovian system, with normal diffusion, which replicates the stationary properties of the reaction-subdiffusion system. Finding a Markovian system such as this allows us to define effective diffusion coefficients and to interpret the behaviour in the original system more easily. We also discuss the conditions under which the transient dynamics of the original system can be approximated by the time-dependent behaviour of this effective Markovian system.

At first it may seem surprising that one should be able to replicate the features of a manifestly non-Markovian system with a Markovian system. However, suppose that, in the long-term, the reaction-subdiffusion system reaches a stationary state. This state may be patterned or homogeneous. Since the non-Markovian `memory' effects in the system are only detectable in dynamic quantities, one might suspect that the stationary behaviour can be described without the use of a non-Markovian memory kernel or a fractional derivative. In this case, one ought to be able to write down a Markovian system which has the same stationary properties as the original non-Markovian system.

We propose a Markovian system of the form
\begin{align}
\frac{\partial \rho_i\left(x,t\right)}{\partial t} &= \frac{\partial^2}{\partial x^2} \left[\hat D_i\left(\boldrho\right)\rho_i \right] +f_i\left(\boldrho\right)\nonumber \\
&= \frac{\partial}{\partial x} \left[D_{i1}\left(\boldrho\right) \frac{\partial\rho_1}{\partial x}+ D_{i2}\left(\boldrho\right) \frac{\partial\rho_2}{\partial x}\right] +f_i\left(\boldrho\right) ,
 \label{effective}
\end{align}
where we define
\begin{align}
 D_{i j}\left(\boldrho\right) \equiv \frac{\partial}{\partial \rho_j} \left[\hat D_i\left(\boldrho\right) \rho_i\left(x\right) \right]. \label{coefficients1}
\end{align}

The specific choice of $\hat D_i\left(\boldrho\right)$, which replicates the stationary properties of the original system, will be described below. The system in Eq. (\ref{effective}) has the same reaction terms as that in Eq.~(\ref{full}), but the anomalous diffusion term has been replaced by a cross-diffusion term. The (effective) diffusion coefficient $\hat D_i$ depends on the local concentration vector $\boldrho(x,t)$. That is, the transport of one substance can be affected by the presence or absence of particles of all types. It is important to note that the diffusion term in Eq.~(\ref{effective}) is not dependent on the history of the system.

To motivate the proposed effective cross-diffusive dynamics and to define the effective diffusion coefficients $\hat D_i$, we first focus on the condition for the instability of a perturbation with wavenumber $k$. In the system described by Eq.~(\ref{effective}) this mode is unstable if 
\begin{align}
&\left[ f_{11} - k^2 D_{11}\left(\boldrho^0\right) \right]\left[f_{22} - k^2 D_{22}\left(\boldrho^0\right) \right] \nonumber \\
&- \left[ f_{12} - k^2 D_{12}\left(\boldrho^0\right) \right]\left[ f_{21} - k^2 D_{21}\left(\boldrho^0\right) \right] < 0 . \label{stabilityconditioneffective}
\end{align}
A detailed derivation of Eq.~(\ref{stabilityconditioneffective}) is given in \cite{Madzvamuse2015, gambinocrossdiffusion}. We define the effective diffusion coefficients for the subdiffusive system by requiring that Eq.~(\ref{stabilityconditioneffective}) is the same condition as Eq.~(\ref{stabilitycondition}). The following choice satisfies this requirement
\begin{align}
\hat D_i\left(\boldrho\right) = \frac{\sigma_i^2}{\eta_i^{\alpha_i}} \left\{\frac{R^-_i\left[\boldrho\left(x \right)\right]}{\rho_i\left(x \right)} \right\}^{1-\alpha_i}. \label{hatcoeff}
\end{align}
Making this choice we obtain
\begin{align}
D_{ij}\left(\boldrho^0 \right) = \frac{\sigma^2_i}{\eta_i^{\alpha_i}}\left[ p_i^{1-\alpha_i}\delta_{ij}+ A_{ij} \rho_i^0 p_i^{-\alpha_i} \left(1-\alpha_i\right)\right]. \label{coefficients2}
\end{align}

By choosing the effective diffusion coefficients in this way, we ensure that the Markovian system and the original subdiffusive system experience the Turing instability for the same sets of system parameters. Furthermore, the same sets of modes are unstable in the Markovian system as in the original subdiffusive system. 

However, having the same sets of unstable modes does not necessarily mean that the two systems converge to the same patterned states in the long-time limit. This is because in order for the steady-state patterns to be produced, the exponential growth of the unstable modes must be curtailed by the non-linearities in the reaction rates. This is not accounted for in the linear stability analysis. That being said, one can indeed show that the stationary patterns produced in the both systems obey the same stationary equation. We first note the following property of the Riemann-Liouville derivative
\begin{align}
{}_0 D_t^{1-\alpha}\left\{ e^{pt}\right\} &= \frac{1}{\Gamma\left(\alpha\right)}\frac{\partial}{\partial t} \int_0^t \frac{e^{p\left(t-\tau\right)}}{ \tau^{1-\alpha}} d\tau \nonumber \\
&= \frac{1}{\Gamma\left(\alpha\right)}\left[t^{\alpha-1} + e^{pt}\int_0^t p \tau^{\alpha-1} e^{-p\tau}dt' \right] \nonumber \\
&= \frac{1}{\Gamma\left(\alpha\right)}\left\{t^{\alpha-1} + e^{pt}p^{1-\alpha}\left[\Gamma\left( \alpha\right) - \Gamma\left(\alpha, pt \right)\right] \right\},
\end{align}
where $\Gamma\left(\alpha, x\right) = \int_x^\infty s^{\alpha-1} e^{-s} ds$ is the upper incomplete gamma function \cite{magnusspecialfunctions}, which has the property $\lim_{x\to\infty} \frac{\Gamma\left(\alpha,x\right)}{x^{\alpha-1}e^{-x}} = 1$. Therefore, for $t \gg \frac{1}{p}$, 
\begin{align}
{}_0 D_t^{1-\alpha}\left\{ e^{pt}\right\} \approx  p^{1-\alpha}e^{pt}. \label{fracderexp}
\end{align}

Using Eq.~(\ref{fracderexp}) in combination with Eq.~(\ref{full}) and the fact that the concentrations do not vary in time in the stationary state, one can deduce that the stationary state in the subdiffusive system obeys
\be
 \frac{\sigma_i^2}{\eta_i^{\alpha_i}}  \frac{\partial^2}{\partial x^2}\left\{ \left[ \frac{R^-_i\left[\boldrho\left(x \right)\right]}{\rho_i\left(x \right)} \right]^{1-\alpha_i}\rho_i\left(x \right) \right\}+ f_i\left[\boldrho\left(x\right)\right]=0, \label{stationary state}
\ee
which, if one notes the definition of $\hat D_i\left(\boldrho\right)$ in Eq.~(\ref{hatcoeff}), is the relation defining the stationary state of Eq.~(\ref{effective}).

\subsection{Critical ratio of the effective coefficients required for pattern formation}\label{section:criticalratio}
So far, we have discussed conditions for the instability of perturbations with specific wavenumbers $k$. We now proceed to establish a condition for the onset of Turing patterns in the subdiffusive system. Such a criterion, a version of which has been written down before in \cite{yadavmilu}, is made far simpler with the use of the effective diffusion coefficients in Eq.~(\ref{coefficients1}).  Since the effective Markovian system has been constructed so that the criterion for the instability of a particular mode $k$ is the same as in the original subdiffusive system, the criterion for Turing pattern formation in general will also be the same in both systems. We therefore work with the effective system.

Turing patterns are formed when a finite range of Fourier modes with non-zero wavenumbers is unstable but the $k=0$ mode is stable, such that Eq.~(\ref{homogeneousstability}) is satisfied. In order to find a condition for the formation of Turing patterns, one identifies the value of $k^2$ for which the left-hand side of Eq.~(\ref{stabilityconditioneffective}) is minimal. We refer to this value as the `critical' Fourier mode. In order for any other mode to satisfy Eq.~(\ref{stabilityconditioneffective}), this critical mode must also do so. One finds
\begin{align}
k^2_{\mathrm{crit}} = \frac{D_{21}f_{12}+D_{12}f_{21}-D_{22}f_{11}- D_{11}f_{22}}{2 \left( D_{12}D_{21} - D_{11}D_{22}\right)} ,
\end{align}
where we keep in mind that $D_{ij} = D_{ij}\left(\boldrho^0\right)$. Substituting this result into Eq.~(\ref{stabilityconditioneffective}), one obtains the following condition for the presence of Turing patterns \cite{Madzvamuse2015, gambinocrossdiffusion}
\begin{widetext}
\begin{align}
\frac{D_{22}}{D_{11}} >\left[\frac{1}{f_{11}}\left(\sqrt{f_{11} f_{22} - f_{12} f_{21}} \pm \sqrt{-\left( f_{12} - \frac{D_{12}}{D_{11}}f_{11}\right) \left(f_{21} - \frac{D_{21}}{D_{11}}f_{11} \right)} \right)\right]^2  - \frac{D_{21}D_{12}}{D_{11}^2}. \label{fullturingcondition}
\end{align}
\end{widetext}

When the effective cross-diffusion terms are zero ($D_{12}=D_{21}=0$) this reduces to the condition for Turing instability in regular reaction-diffusion systems \cite{crossbook},
\begin{align}
\frac{D_{22}}{D_{11}} > \left[\frac{1}{f_1}\left(\sqrt{f_{11}f_{22} - f_{12}f_{21}} \pm \sqrt{- f_{12}f_{21}} \right) \right]^2.\label{nocross}
\end{align}
This reduction to a known result for diffusive systems underlines the fact that the effective diffusion coefficients encapsulate the net diffusivity in the reaction-subdiffusion system.

For the cross-terms to be zero, the death rates of both species must be independent of the concentration of the other, or the diffusion of both species must be Markovian. The choice of whether the criteria Eq.~(\ref{fullturingcondition}) and Eq.~(\ref{nocross}) contain a plus or a minus sign is determined by which of species 1 or 2 is the activator/inhibitor. This can be seen realising that the critical ratio of the inhibitor diffusion constant to that of the activator for Turing pattern formation to occur should reduce to a number which is greater than unity in the normally diffusing case \cite{yadavmilu, murrayII}. If species 1 is the activator, then Eq.~(\ref{fullturingcondition}) and Eq.~(\ref{nocross}) should contain a plus sign.

\subsection{Limit of large removal rates}\label{subsection:largedeathrate}
Naturally, the Markovian and subdiffusive systems described by Eqs.~(\ref{full}) and (\ref{effective}) do not, in general, have identical dynamics. However, one can show that the dynamics of the subdiffusive system are replicated to a good approximation in the corresponding effective Markovian system, when the particle removal rates are large. In this limit, particles survive for only short amounts of time on average, so the time horizon for memory effects is also small and, consequently, the dynamics can be approximated as memoryless. 

To show this, let $p_i\left(x,t\right) = \frac{R_i^-\left[\boldrho\left(x,t\right)\right]}{\rho_i\left(x,t\right)}$ be the per capita removal rate for particles of type $i$. When the death rate is sufficiently large, one obtains the following
\begin{align}
{}_0D^{1-\alpha_i}_t &\left\{ \rho_i\left(x, t\right) e^{ \int_0^t p_i\left(x,t'\right) dt'}\right\} \nonumber \\
&\approx  \rho_i\left(x,t\right) p_i\left(x,t \right)^{1-\alpha_i} e^{ \int_0^t p_i\left(x,t'\right) dt'}. \label{approxresult2}
\end{align}
We refer the reader to Appendix \ref{appendix:fastdeath} for further details of this approximation.
 Noting that $\frac{\sigma_i^2}{\eta_i^{\alpha_i}} p_i\left(x,t \right)^{1-\alpha_i} = \hat D_i\left[ \boldrho\left(x,t\right)\right]$, the right-hand sides of Eqs.~(\ref{full}) and (\ref{effective}) are found to be approximately equal in the regime in which the approximation Eq.~(\ref{approxresult2}) is valid.\\

\subsection{Interpretation}\label{subsection:interp}
We have shown that an appropriately constructed normally diffusive system can replicate some of the most important features of the original subdiffusive system, in particular the location of any Turing instability in parameter space, and the resulting patterns. This has allowed us to define effective diffusion coefficients for the subdiffusive system. In this section, we provide further insight as to why this is possible.

Before we do this however, let us first contrast the effective diffusion coefficients $\hat D_i(\boldrho)$ with the quantities $D_{ij}(\boldrho)$ [see Eq. (\ref{coefficients1})]. The coefficients $\hat D_i\left(\boldrho\right)$ appear inside the second-order spatial derivative in Eq.~(\ref{effective}); they represent the proclivity of particles to hop away from a given location. The coefficient $D_{ij}\left(\boldrho\right)$ on the other hand describes how the transport of particles of type $i$ is affected by a local gradient of the concentration of particles of type $j$. Which set of coefficients one decides to use is a matter of taste. The coefficients $D_{ij}\left(\boldrho\right)$ have the advantage of highlighting the cross-diffusive nature of the system while the $\hat D_i\left(\boldrho\right)$ are more succinct. The $D_{ij}\left(\boldrho\right)$ are also useful for writing down the criterion for Turing instability, as was demonstrated in Section \ref{section:criticalratio}.

So far, we have introduced effective diffusion coefficients for subdiffusive systems in the context of Eq.~(\ref{effective}), but we have not related them to the statistics of particle trajectories. We would now like to provide further interpretation, and establish how the well-known diffusive law $\langle x^2 \rangle = 2 D t$ can be obtained in the context of subdiffusive systems. As we will see this can be achieved by defining a suitable ensemble of particles, and the connection with the effective diffusion coefficients can be made.

To do this, we consider the following equation,
\begin{align}
\frac{\partial P_i\left(x,t\right)}{\partial t} &= \frac{\sigma^2_i}{\eta_i^{\alpha_i}}\frac{\partial^2}{\partial x^2}\left\{ e^{- p_i t}  {}_0 D_t^{1-\alpha_i} \left[ e^{p_i t} P_i\left(x,t \right)\right]\right\} \nonumber \\
& - \theta p_iP_i\left(x,t\right) . \label{effectiveequation}
\end{align}
in which $\theta$ is a parameter which may take values $0$ or $1$. Eq.~(\ref{effectiveequation}) encapsulates the subdiffusive terms of Eq. (\ref{full}), and, for $\theta=1$, the annihilation reactions when the removal rate $p_i$ is constant and uniform (as it would be the case in the homogeneous steady state). In the original system, subdiffusive motion and removal are the only processes an individual particle can undergo once it has been created. We assume that the initial condition at time $t=0$ is given by $P_i(x,t=0)=\delta(x)$. Let us discuss possible microscopic processes described by Eq.~(\ref{effectiveequation}).  

For $\theta = 1$, the equation describes a system in which particles subdiffuse by drawing waiting times from the Mittag-Leffler distribution, and in which they are also subject to a constant removal rate $p_i$. The solution $P_i\left(x,t\right)$ of Eq. (\ref{effectiveequation}) describes the density of particles of type $i$ present at time $t$ at position $x$. The total number of particles in the system, $\int P_i\left(x,t\right) dx =e^{-p_i t}$, decreases with time. We define $C_i\left(x,t\right) \equiv P_i\left(x,t\right)/[\int P_i\left(x',t\right) dx']$; this is the probability density for the position of {\em surviving} particles, that is particles still present in the system at time $t$.  

The subdiffusive law is obtained by examining the mean squared displacement of these surviving particles. To see this we define $\avg{x(t)^2}_{\rm surv}=\int C_i(x,t) x^2 dx$. One then finds
\begin{align}
\avg{x(t)^2}_{\rm surv} = 2\frac{\sigma^2_i}{\eta_i^\alpha} \frac{1}{\Gamma\left( 1 + \alpha_i\right)} t^{\alpha_i}.
 \label{subdiffusivelaw}
\end{align}
 Eq. (\ref{subdiffusivelaw}) is derived in Appendix \ref{appendix:effectiveprocess}.

We now turn to the the case $\theta=0$, which can be obtained from the scenario for $\theta=1$ by adding a term $p_iP_i(x,t)$ on the right-hand side of Eq. (\ref{effectiveequation}). This indicates additional particle production. More precisely, one representation of the case $\theta=0$ is a dynamics in which any particle that is removed is immediately replaced by another identical particle, which draws a new waiting time. This reflects the behaviour in the stationary state of the full system in Eq. (\ref{generalreactdiff}). In the stationary state the particle number at each location is constant, and $R_i^+(\boldrho^0)=R_i^-(\boldrho^0)$. The per capita removal rate of particles of type $i$ is given by $p_i=R_i^-(\boldrho^0)/\rho_i^0$, and removal and production balance. 
The density $P_i(x,t)$ remains normalised at all times for $\theta=0$, and we define $\avg{x(t)^2}_{\rm replace}=\int P_i(x,t) x^2 dx$. It is this ensemble which leads to diffusive behaviour, 
\begin{align}
\avg{x(t)^2}_{\rm replace}\approx 2\frac{\sigma^2_i}{\eta_i^\alpha} p_i^{1-\alpha_i} t . \label{diffusivelaw}
\end{align}
This relation is again derived in Appendix \ref{appendix:effectiveprocess}.

From Eq.~(\ref{diffusivelaw}), one can read off the effective diffusion coefficient $\hat D_i \left( \boldrho^0\right)=\frac{\sigma^2_i}{\eta_i^\alpha} p_i^{1-\alpha_i}$. This coincides with the earlier definition in Eq. (\ref{hatcoeff}).

In summary, if one includes the removal term in Eq.~(\ref{effectiveequation}) ($\theta = 1$), one finds that the ensemble of {\em surviving} particles generates subdiffusive statistics. If this term is not present (i.e. removed particles are replaced), standard diffusive statistics $\avg{x(t)^2}\propto t$ ensue, despite the non-standard transport term in Eq.~(\ref{effectiveequation})

\section{Verification using numerical integration}\label{section:simulations}
\subsection{Lengyel-Epstein model}\label{sec:lengyelmain}
To test the conclusions of Section \ref{section:effectivesystem}, we numerically integrate the reaction-subdiffusion-equation (\ref{full}) and the corresponding normally diffusive system in Eq. (\ref{effective}) and then compare the numerical solutions. We do this for the example of the Lengyel-Epstein model \cite{lengyelepstein1}, which was introduced primarily as a way of modelling the $\textrm{ClO}_2^-$--$\textrm{I}^-$--MA reaction. This chemical system exhibits Turing patterns experimentally \cite{castets, lengyelepstein2}. We use a simplified two-species version of the full model, and focus on the case of one spatial dimension. The system involves two types of particles, labelled $A$ and $B$, which each undergo a (sub)diffusion process, potentially with different anomalous exponents. Particles at the same location can also react. The reactions are described by the terms
\begin{align}
f_{A}\left(\boldrho\right) &= aN - b \rho_A - 4 \frac{cN \rho_A \rho_B}{d N^2 + \rho_A^2} , \nonumber \\
f_{B}\left(\boldrho\right) &= b \rho_A - \frac{cN \rho_A \rho_B}{d N^2 + \rho_A^2}, \label{le1}
\end{align}
where $\rho_A$ and $\rho_B$ are the particle concentrations, and where $a, b, c, d$ and $N$ are positive model parameters.
The first term in the definition of $f_A$ describes the production of particles of type $A$, and the first term in $f_B$ represents the production of $B$-particles. The remaining terms describe removal processes; the corresponding per capita death rates for the two species are given by
\begin{align}
\frac{R_A^-}{\rho_A} &= b + 4 \frac{cN \rho_B}{d N^2 + \rho_A^2} , \nonumber \\
\frac{R_B^-}{\rho_B} &= \frac{cN \rho_B}{d N^2 + \rho_A^2} . \label{le2}
\end{align}
The Lengyel-Epstein system has a homogeneous deterministic fixed point at $\rho_A^0 = \frac{aN}{5b}, \rho_B^0 = \frac{bdN}{c}\left[1+\left(\frac{\rho_A^0}{\sqrt{d}N}\right)^2 \right]$. Using Eq.~(\ref{homogeneousstability}), the homogeneous fixed point in the well-mixed system is stable so long as $ca~>~\frac{3}{5}a^2~-~25b^2d$.

\subsection{Numerical integration method}

The numerical integration of Eq.~(\ref{full}) is made difficult by the fact that the evolution of the system depends not only on the concentrations at the present time, but also on the concentrations at all earlier times. This is due to the nature of the fractional derivative ${}_0D_t^{1-\alpha}$, defined in Eq.~(\ref{fracdiffdef}). Further complications arise from the coupling of reaction and subdiffusion terms, {\em viz.} the presence of the exponential term inside the fractional derivative in Eq. (\ref{full}).  To our knowledge, most existing numerical methods for integrating reaction-subdiffusion equations have not taken this coupling into account \cite{fracnumeric1, fracnumeric2, fracnumeric3}. 

The key to integrating Eq.~(\ref{full}) is in recognising that the fractional derivative can be expressed in discretised form using the equivalence of the Gr\"{u}nwald-Letnikov and the Riemann-Liouville definitions of the fractional derivative, which is valid when $f\left(t\right)$ is continuous and $f'\left(t\right)$ is integrable \cite{podlubnybook,scherernumerical},
\begin{align}
{}_0D_t^{1-\alpha} f\left(t\right)= \frac{1}{\left(\Delta t\right)^\alpha} \sum_{j=0}^{\left[\frac{t}{\Delta t}\right]} \left( -1\right)^j \binom{\alpha}{j} f\left(t - j\Delta t\right)+{\cal O}(\Delta t), \label{glderiv}
\end{align}
where $\left[x\right]$ denotes the integer part of $x$ and where $\binom{\alpha}{j}$ is the generalised binomial coefficient
\begin{align}
 \left(\begin{array}{c} \alpha \\ j \end{array}\right) = \frac{\alpha \left(\alpha-1\right) \left( \alpha - 2\right) \cdots \left(\alpha-j+1\right)}{j!} .
\end{align}
 \begin{figure*}[t!!]
\includegraphics[width=0.47\textwidth]{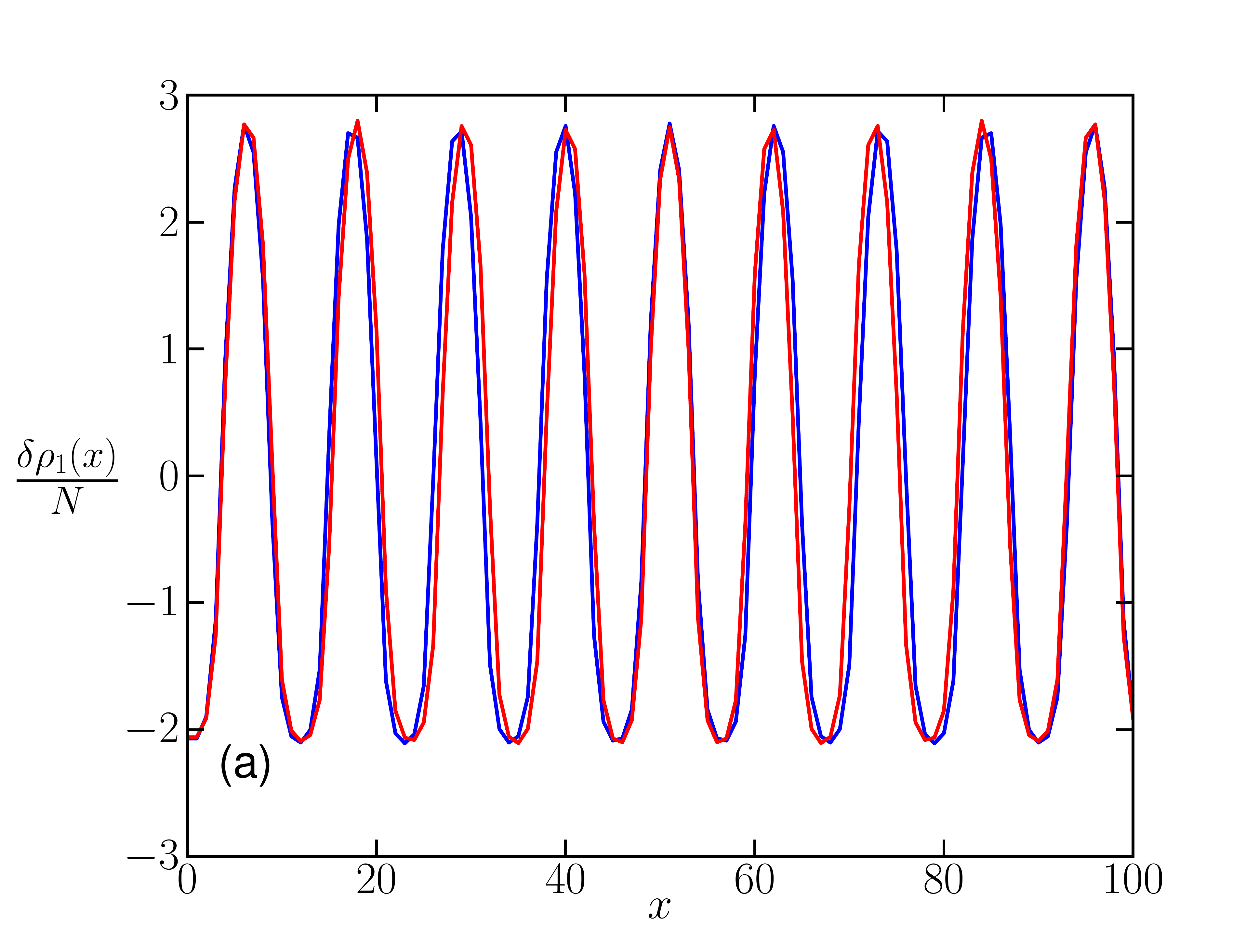}
\includegraphics[width=0.47\textwidth]{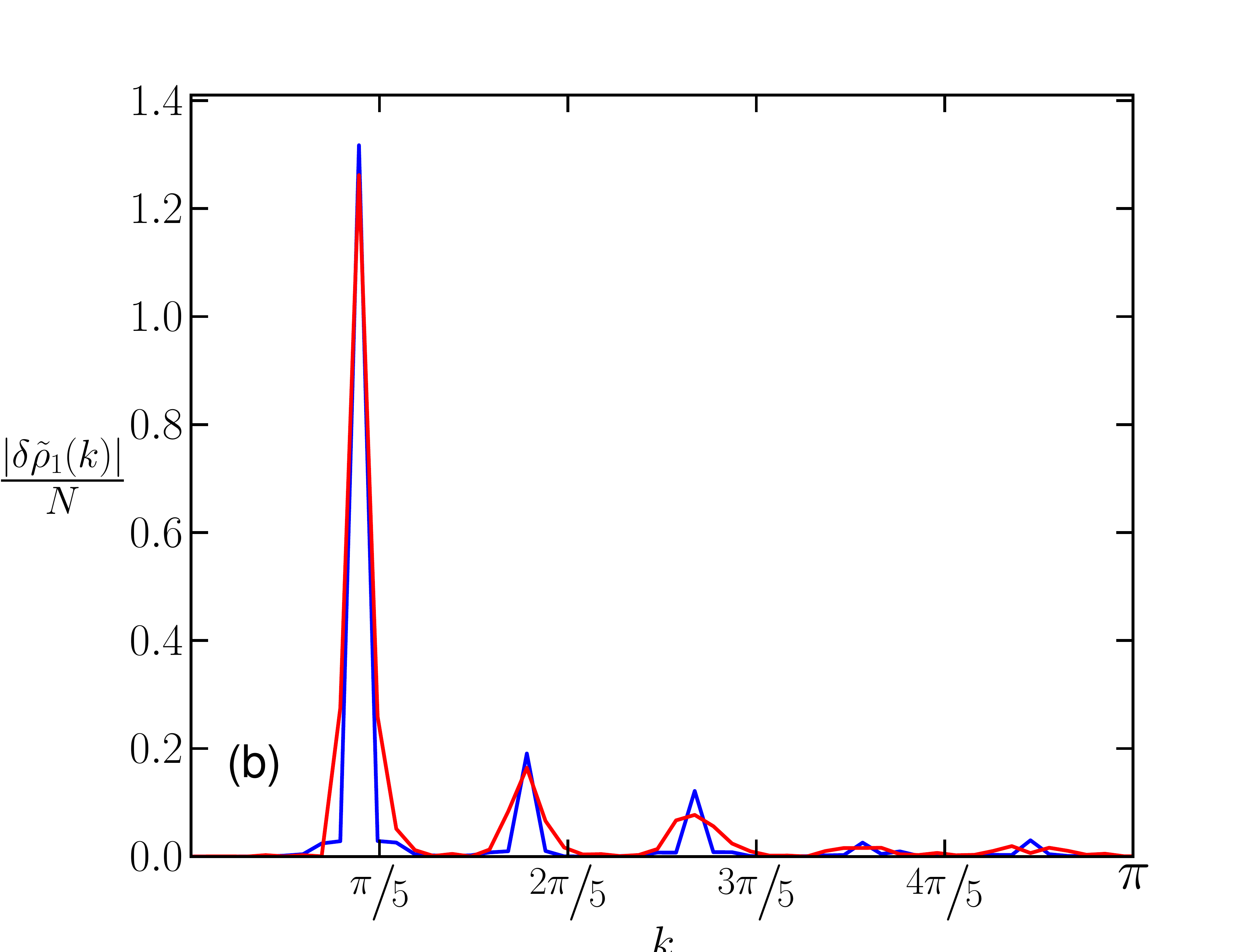}
\vspace{0.5em}
	\caption{ (a) Comparison of patterns obtained from numerically integrating the full reaction-subdiffusion equation (\ref{full}), shown as blue line, and the effective Markovian system in Eq.~(\ref{effective}), shown as red line. Data shows deviations from the homogeneous fixed point for species A of the the Lengyel-Epstein model (c.f. Sec. \ref{sec:lengyelmain}). (b) Fourier spectra of these patterns. The model parameters are $a = 2$, $b = 0.13$, $c = 1$, $d = 1$, $N = 1000$,  $\alpha_A = 0.5$, $\alpha_B = 1$, $\frac{\sigma_A^2}{\eta_A^{\alpha_A}} = 1.05$, $\frac{\sigma_B^2}{\eta_B^{\alpha_B}} = 0.33$, $p_A = 0.65$, $p_B = 0.294$. Both systems were run until $t = 300$ so that the patterns have sufficient time to form. The cut-off time used was $t_{\mathrm{cut}} = 10$.  } 
	\label{fig:stationarypatterns}
\end{figure*}
The numerical approach is then based on first-order Euler-forward integration. We also discretise space and operate on a one-dimensional lattice with spacing $\Delta x$ and periodic boundary conditions. The Laplacian in one dimension, $\frac{\partial^2 f}{\partial x^2}(x)$, is discretised as $[f(x-\Delta x)-2f(x)+f(x+\Delta x)]/\Delta x^2$. At each time-step the fractional derivatives ${}_0 D_t^{1-\alpha} \left( \exp\left\{ \int_0^t \frac{R^-_i \left[\boldrho\left(x,t' \right)\right]}{\rho_i\left(x,t' \right)} dt'\right\} \rho_i\left(x,t \right)\right)$ must be evaluated for every site on the lattice. This means that it is necessary to keep a history of the concentrations for every point on the lattice. For efficiency, it is also convenient to keep the history of the integral that appears in the exponential.\\
The `short-memory' principle \cite{podlubnybook} allows us to simplify matters and speed up the computation. The definition of the fractional derivative in Eq.~(\ref{fracdiffdef}) indicates that times closer to the present contribute more than times further in the past. Using a `cut-off' $t_{\mathrm{cut}}$ in how far back one goes in time to calculate the fractional derivative ensures that the numerical evaluation does not slow to halt as $t$ becomes large. One must be careful however to choose a value of $t_{\mathrm{cut}}$ which does not interfere appreciably with the results. Further details of the method are discussed in Appendix \ref{numericalmethod}.

The numerical integration is carried out on a lattice with $101$ nodes and with spacing $\Delta x= 1$, which is much smaller than the wavelength of the patterns produced in the examples. The time step used in the Euler-forward integration is $\Delta t=10^{-4}$. Further model parameters are given in the figure captions.
 
\subsection{Numerical comparison of subdiffusive and corresponding normally diffusive systems}
\subsubsection{Stationary patterns}

\begin{figure*}[t!!]
	\centering
        \includegraphics[width=0.48\textwidth]{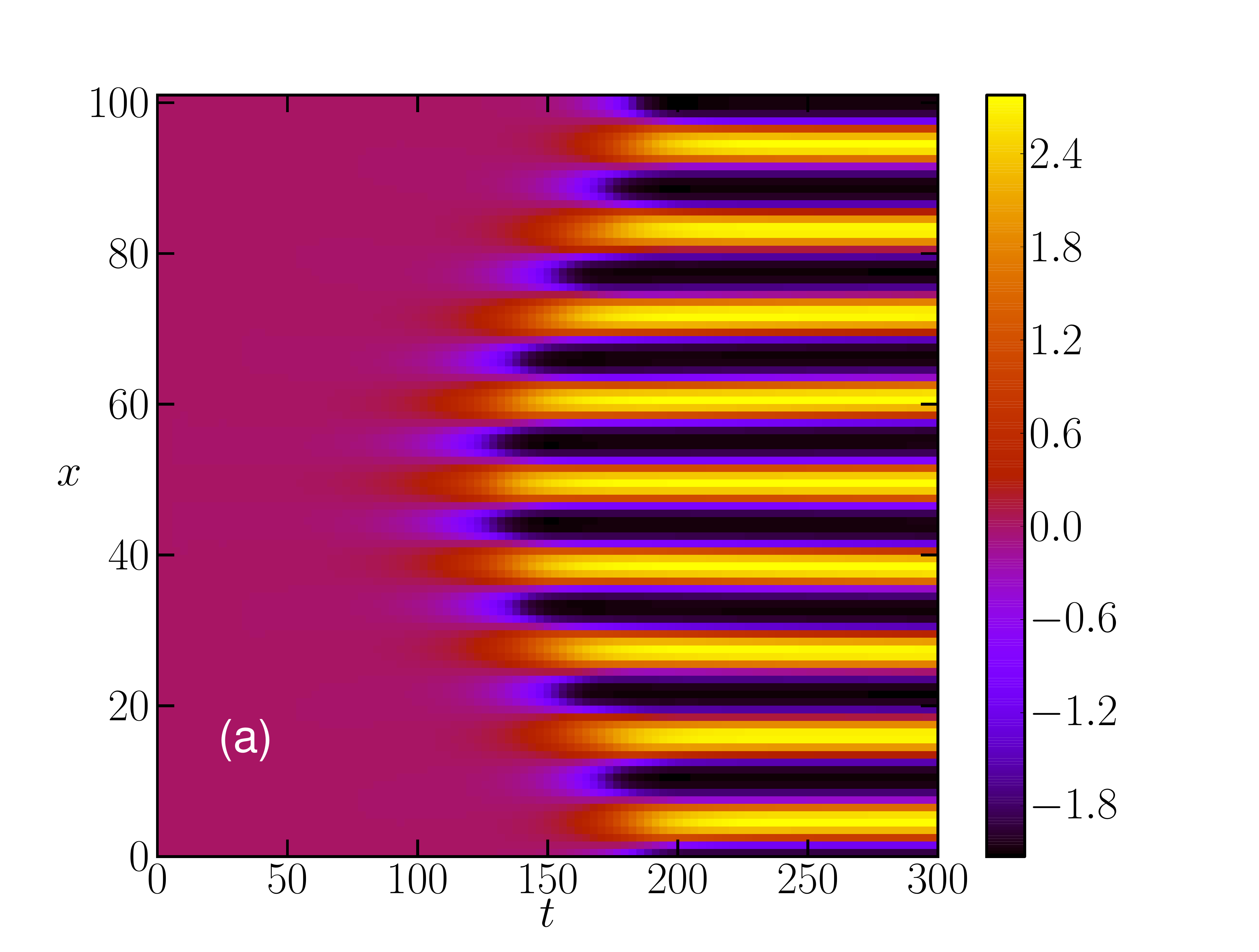}
		\includegraphics[width=0.48\textwidth]{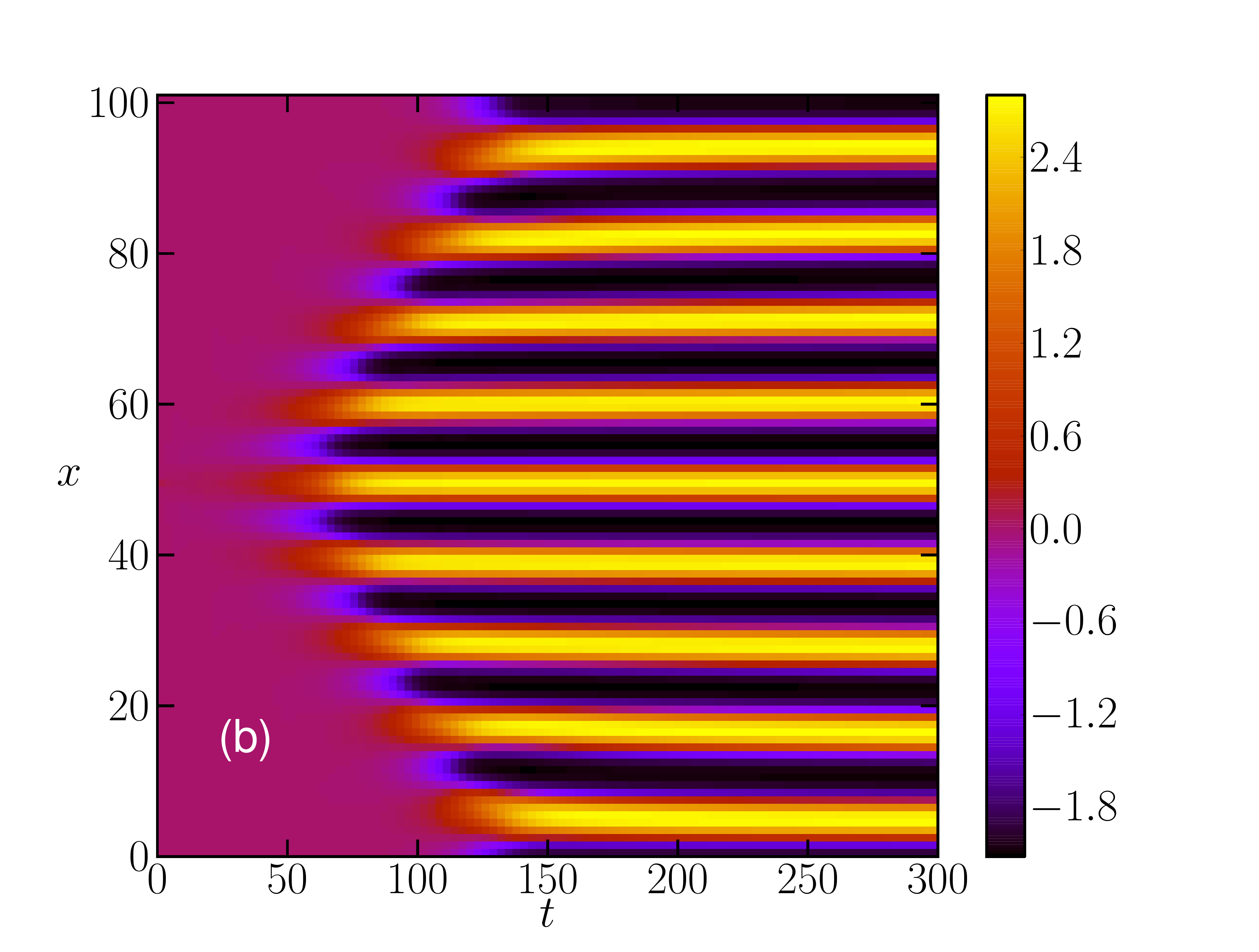}
	\caption{Comparison of the transient approach to the patterned state in the (a) original and (b) effective Markovian systems for the same model and parameters as in Fig. \ref{fig:stationarypatterns}. We show $\delta \rho_A/N$ as relief. The patterns are fully formed at roughly $t = 150$ in the effective system but take until around $t = 200$ to form in the subdiffusive system. The system is initialised at the homogeneous fixed point, with a deviation of magnitude $10$ at $x=51$ at $t=0$. The cut-off time used was $t_{\mathrm{cut}} = 10$.}
	\label{fig:slowdeath}
\end{figure*}

As discussed in Section \ref{section:deffeff}, for fixed model parameters we expect the same stationary patterns to result from the reaction-subdiffusion system and the corresponding effective Markovian system in the long-term.  The patterns in the Lengyel-Epstein model shown in the left-hand panel of Fig. \ref{fig:stationarypatterns} confirm this; the amplitude, periodicities and general shapes of the patterns in the original and the effective systems match. The similarity of the patterns is further evidenced in the Fourier spectra shown in the right-hand panel. We attribute the remaining differences to inaccuracies in the numerical methods.
Fig. \ref{fig:slowdeath} shows that the transient approach to the patterned stationary state in the subdiffusive system differs from that in the effective cross-diffusive system.

\subsubsection{Time-evolution for large removal rates}
\begin{figure*}[t!!]
\includegraphics[width=0.47\textwidth]{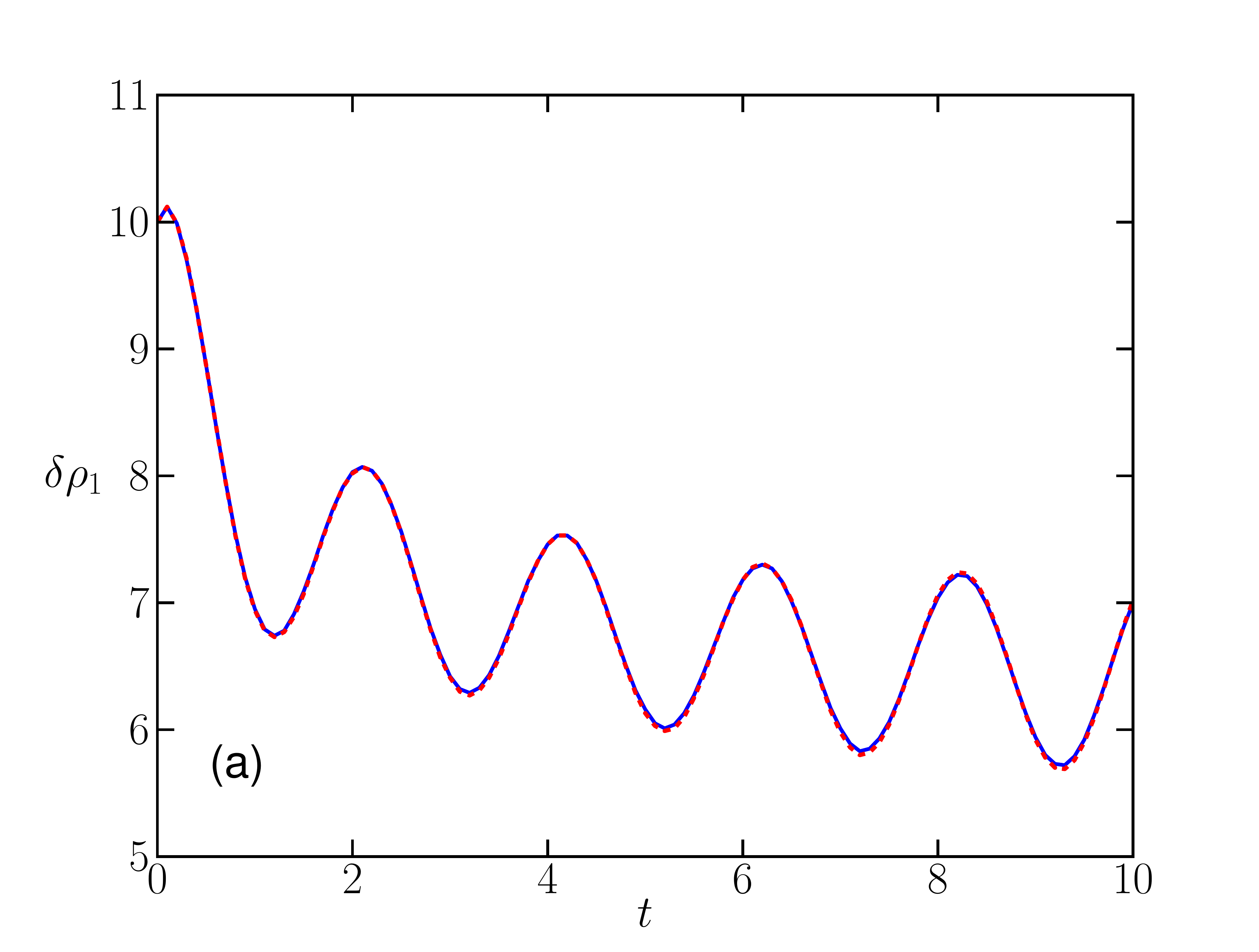}
\includegraphics[width=0.47\textwidth]{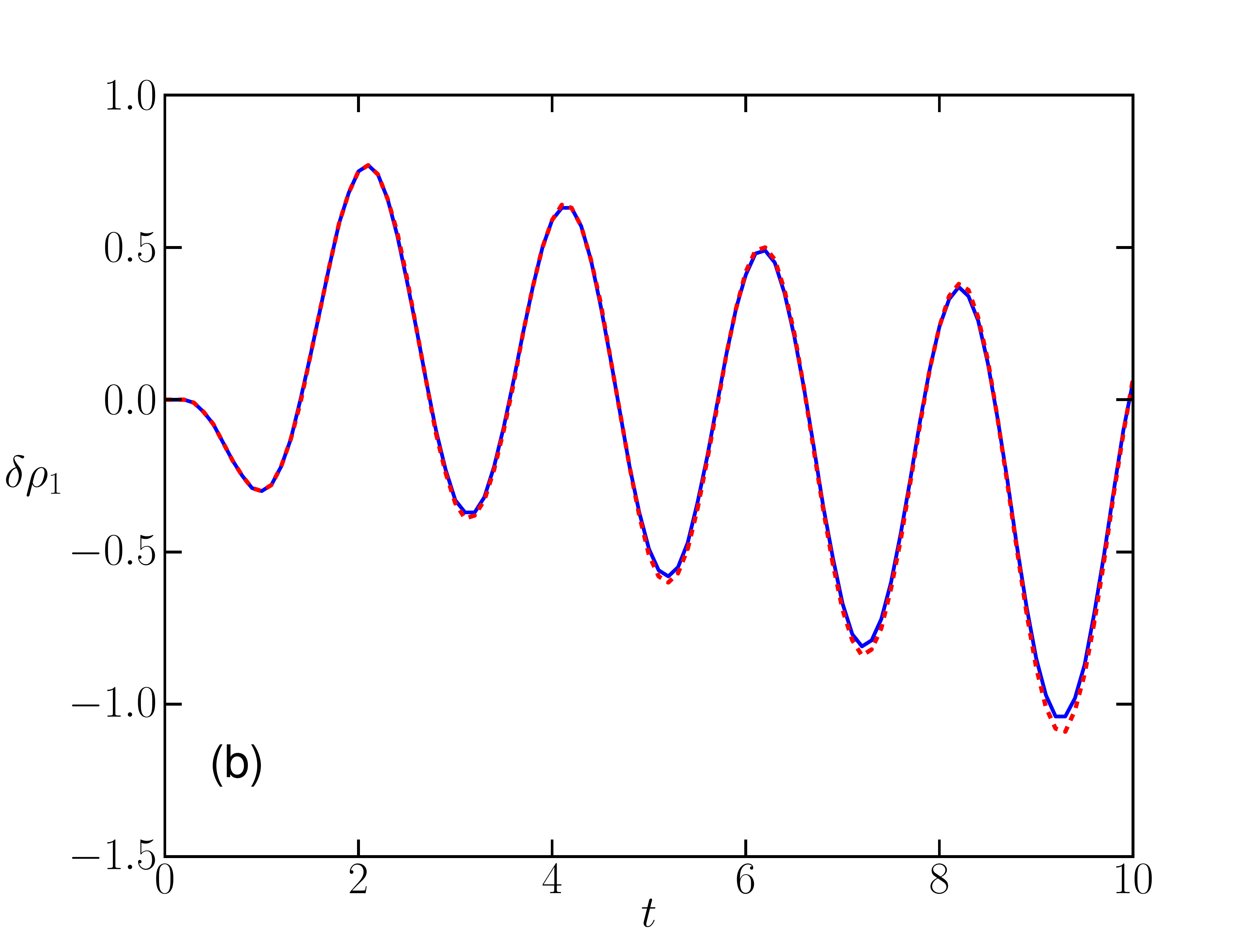}
\caption{Comparison of the initial transient behaviour in the original and effective Markovian systems for the case of large removal rate for the subdiffusing component. Integration is initialised at the homogeneous fixed point but with a deviation of magnitude $10$ at $x=0$ at $t=0$. (a) Time evolution of species A at $x=0$ in the reaction-subdiffusion system (blue line) and the corresponding Markovian system (red dashed line). (b) Time-evolution of species A at $x=6$. The model parameters are $a = 27$, $b = 4$, $c = 1$, $d = 1$, $N = 1000$, $\alpha_A = 0.5$, $\alpha_B = 1$, $\frac{\sigma_A^2}{\eta_A^{\alpha_A}} = 0.075$, $\frac{\sigma_B^2}{\eta_B^{\alpha_B}} = 16.7$. The death rate of the subdiffusing species $A$ is $p_A = 20$, the death rate for the normally diffusing species $B$ is $p_B = 0.478$. The cut-off time used was $t_{\mathrm{cut}} = 1$.}
	\label{fig:fastdeath}
\end{figure*}

\begin{figure*}[t!!]
	\centering

        \includegraphics[width=0.47\textwidth]{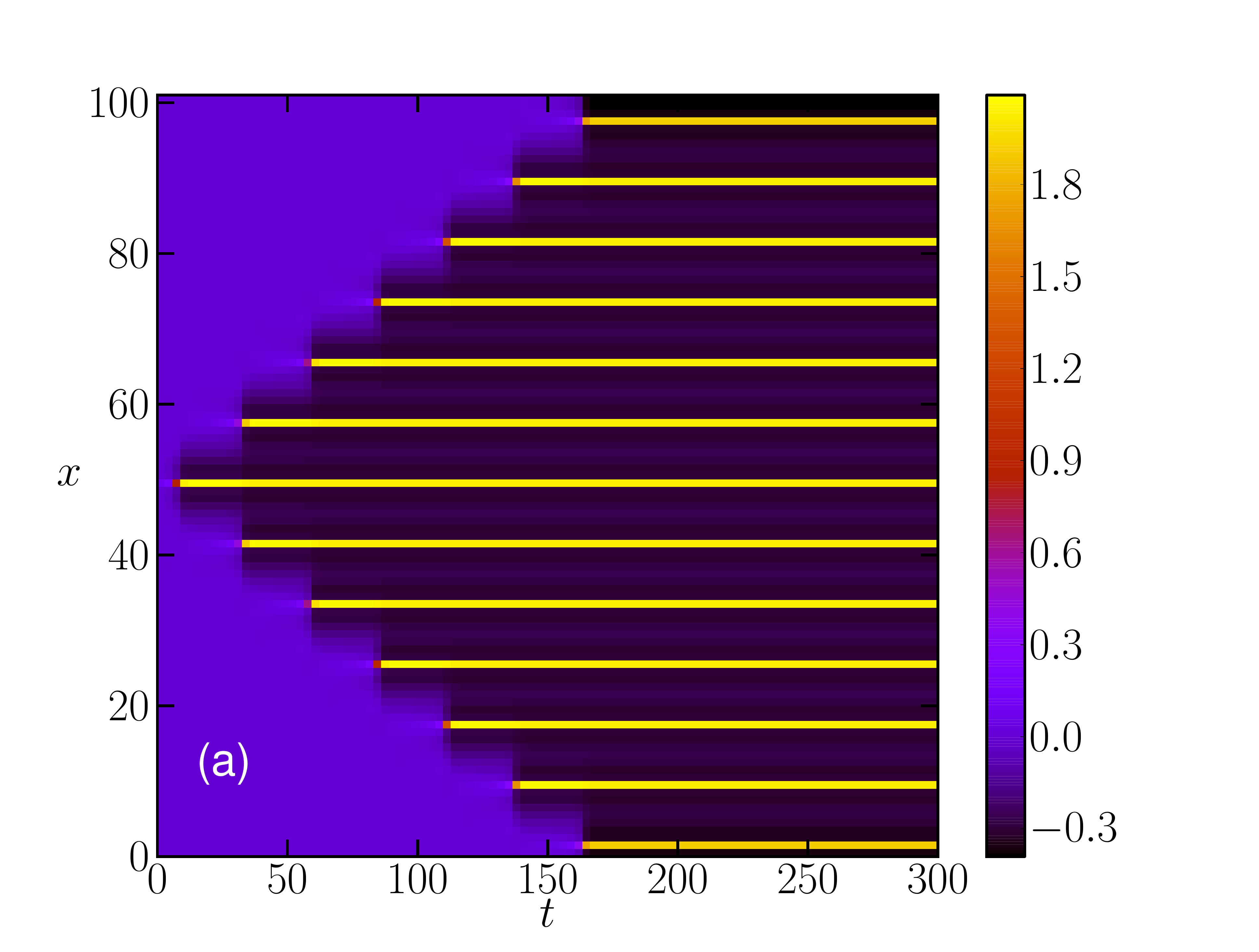}
		\includegraphics[width=0.47\textwidth]{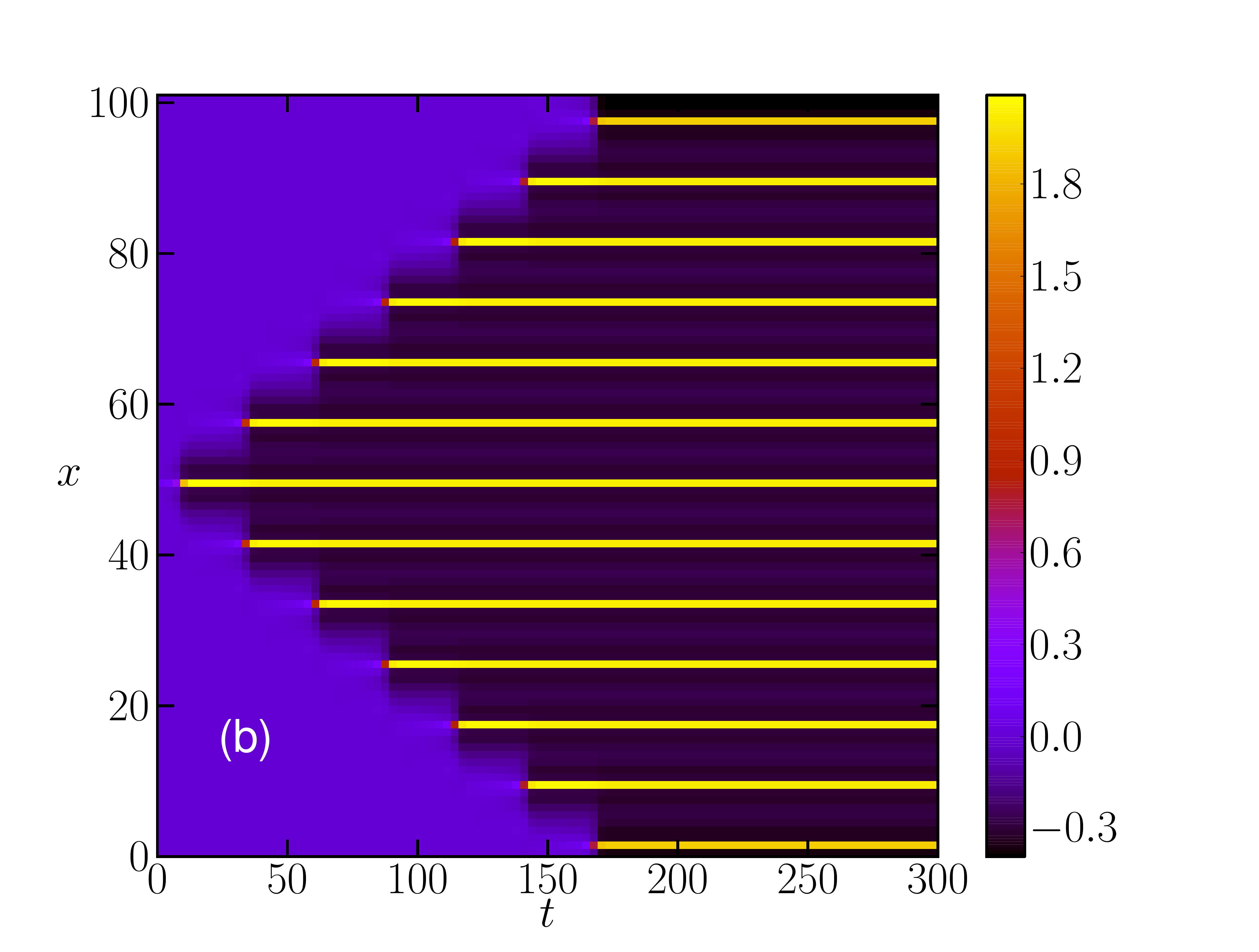}

	\caption{Comparison of the intermediate-time behaviour in the (a) effective Markovian and (b) original systems for the case of large removal rate for the subdiffusing component (see text for details). We show $\delta \rho_A/N$ as relief. The system is initialised at the homogeneous fixed point, with a deviation of magnitude $100$ at $x=51$ at $t=0$. Model parameters are $a = 27$, $b = 4.06$, $c = 1$, $d = 1$, $N = 1000$,  $\alpha_A = 0.5$, $\alpha_B = 1$, $\frac{\sigma_A^2}{\eta_A^{\alpha_A}} = 0.067$, $\frac{\sigma_B^2}{\eta_B^{\alpha_B}} = 33.3$. The death rate of the subdiffusing species is $p_A = 20$, the death rate for the normally diffusing species is $p_B = 0.478$. The cut-off time used was $t_{\mathrm{cut}} = 10$. }
	\label{fig:fastdeath2}
\end{figure*}

We also deduced in Section \ref{subsection:largedeathrate} that the original subdiffusive system and the corresponding effective Markovian system have approximately the same dynamics when the particle removal rates are large. In Figs. \ref{fig:fastdeath} and \ref{fig:fastdeath2}, we verify this for the case in which one species subdiffuses and the other diffuses normally. If one component diffuses normally ($\alpha_i = 1$), the evolution equation for this component is identical in both systems. So we only require the subdiffusing component to have a large removal rate in order for the dynamics to be the same in both systems.\\
Fig. \ref{fig:fastdeath} demonstrates the accuracy of the approximation for large removal rates. We compare the transient time-evolution of the concentration for the original system and for the effective Markovian system. This is done for two different locations in the spatial domain. Fig. \ref{fig:fastdeath2} shows the agreement between the systems over longer time scales as patterns begin to emerge.

\section{Summary and discussion}\label{section:discussion}

In a Markovian reaction-diffusion system, the quantification of the motility of a particular species is comparatively straightforward since one can easily define the diffusion coefficient, which can be related to the rate of change with time of the mean squared displacement for an individual particle. The situation in subdiffusive systems is more complicated. In this paper we have defined the effective diffusion coefficient for a reaction-subdiffusion system, and we are now able to compare the motilities of particle species. These effective diffusion coefficients are dependent not only on the anomalous exponents and the typical waiting times of particles, but also on the reaction rates. This in turn gives rise to cross-diffusive behaviour. Moreover, we have shown that, using the effective diffusion coefficients, the condition for Turing pattern formation in a subdiffusive system can be written in the same form as it would appear in a Markovian system. We have therefore demonstrated that the effective diffusion coefficients are indeed the pertinent quantities with which one can characterise transport effects in a reaction-subdiffusion system. 

We emphasize that while the effective Markovian system replicates the behaviour in the subdiffusive system in the stationary state, it reproduces the transient dynamical behaviour only under special circumstances. Interesting time-dependent phenomena which are peculiar to reaction-subdiffusion systems, such as the failure of front-propagation \cite{sokolov3}, are not necessarily reproduced fully by the effective system. That being said, we note that the concentration profiles in \cite{sokolov1}, which were reported to have unique character due to subdiffusion, can also be produced using an effective Markovian system of the type discussed in this paper. This can be seen through the fact that Eq.~(11) in \cite{sokolov1} is a special case of our Eq.~(\ref{stationary state}) for the particular reaction scheme used in that paper.

In Sec. \ref{subsection:interp} we showed that the mean squared displacement of particles can be thought of as increasing linearly with time even in a subdiffusive system. This view can be taken if one interprets diffusion as a phenomenon defined by an ensemble of particles, rather than by tracking the motion of individual particles. Different choices for the ensemble can lead to subdiffusive or regular diffusive behaviour. In the latter case, the diffusion coefficient, defined from the ensemble view, coincides with the one we obtained from analysing the Turing instability of the system with anomalous diffusion. 

There exist many explanations for the observed cross-diffusive effects in real-life systems. In ecological systems for example, predators pursue their prey and prey avoid their predators \cite{dubey, biktashev}. Mechanisms leading to cross-diffusive behaviour in physical and chemical systems include electrostatic interactions, excluded-volume effects and complexation  \cite{vanag}. Our analysis suggests another possible mechanism: If a system exhibits memory effects due to anomalous transport, cross-diffusion can arise when the removal rate of one species of particle depends on the concentration of another species.

A common way of measuring diffusion coefficients in chemical systems experimentally is the Taylor method, which involves monitoring the spread of the concentration of a drop placed in a laminar flow. A Gaussian profile is then fitted to the data in order to estimate the diffusion coefficients \cite{Taylor, Price}. In such an experiment, individual particles are not tracked and so the diffusive properties are inferred based on macroscopic statistical behaviour. Our work would suggest that cross-diffusive coefficients arising in experiments of this kind could possibly originate from, or at least be affected by, non-Markovian transport.

While we restricted most of the discussion to the case of subdiffusion with Mittag-Leffler distributed waiting times, we note that an effective normally diffusive system can be found for general non-Markovian systems described by Eq.~(\ref{generalreactdiff}); the calculations that were carried out starting from Eq.~(\ref{full}) could just as well be performed with Eq.~(\ref{generalreactdiff}). So effective diffusion coefficients can also be found for reaction-diffusion systems with other types of non-standard diffusion.
\\

{\em Acknowledgements.} JWB thanks the Engineering and Physical Sciences Research Council (EPSRC) for funding (PhD studentship, EP/N509565/1). We would also like to thank Francisco Herrar\'ias-Azcu\'e for his helpful comments.

\begin{appendix}
\onecolumngrid
\section{Criterion for Turing patterns}
\subsection{Derivation of the criterion for Turing instability}\label{appendix:stabilitycriterion}
In this Appendix we derive the condition in Eq. (\ref{stabilitycondition}) for the instability of a Fourier mode with wavenumber $k$. The calculation differs from that in \cite{yadav} in that we avoid the use of contour integration; we also consider the case where both species can subdiffuse. 

We start from the linearised dynamics in Eq.~(\ref{linearisedfrac}), and proceed using the ansatz
\begin{align}
\delta \underline{\tilde\rho}\left(k,t\right) = e^{\lambda_k t} \delta \underline{\tilde\rho}\left(k,0\right) ,\label{ansatz}
\end{align}
where $\lambda_k>0$ is real.  Inserting this into Eq.~(\ref{linearisedfrac}) and using Eq.~(\ref{fracderexp}), we find
 
\begin{align}
\lambda_k \delta\tilde\rho_i\left(k,t\right) &= -k^2 \frac{\sigma^2_i}{\eta_i^{\alpha_i}}\Bigg[ \left(p_i+ \lambda_k\right)^{1-\alpha_i}\delta \tilde\rho_i\left(k,t\right) - \sum_j A_{ij}\rho_i^0   \frac{   p_i^{1-\alpha_i} - \left(\lambda_k + p_i \right)^{1-\alpha_i} }{\lambda_k} \delta \tilde\rho_j\left(k,t\right)\Bigg] + \sum_jf_{ij}\delta\tilde\rho_j\left(k,t\right) , \label{eigen1}
\end{align}
at long times. Eq.~(\ref{eigen1}) can be rewritten in matrix form
\begin{align}
\underline{\underline{M_k}} \delta \underline{\tilde\rho_k} = \underline{0} .\label{matrixform}
\end{align}
In order for the solution to be non-trivial, the matrix $\underline{\underline{M_k}} $ must be singular, that is
\begin{align}
\Delta_k\left(\lambda_k\right) \equiv \mathrm{det} \underline{\underline{M_k}}  = 0. \label{characteristiceq}
\end{align}
Eq.~(\ref{characteristiceq}) determines the possible values that $\lambda_k$ may take. The determinant $ \Delta_k\left(\lambda_k\right)$ can be written as
\BE
\Delta_k\left(\lambda_k\right) &=& \left[\lambda_k +k^2\frac{\sigma^2_1}{\eta_1^{\alpha_1}}\left( \left(\lambda_k + p_1 \right)^{1-\alpha_1}+ A_{11} \rho_1^0  \frac{ \left( \lambda_k + p_1\right)^{1-\alpha_1} - p_1^{1-\alpha_1}}{\lambda_k} \right) - f_{11}\right]\nonumber \\
&&\times\left[\lambda_k +k^2\frac{\sigma^2_2}{\eta_2^{\alpha_2}}\left( \left(\lambda_k + p_2 \right)^{1-\alpha_2}+ A_{22} \rho_2^0  \frac{ \left( \lambda_k + p_2\right)^{1-\alpha_2}-p_2^{1-\alpha_2}}{\lambda_k} \right)-f_{22}\right] \nonumber \\
&&-\left[k^2\frac{\sigma^2_2}{\eta_2^{\alpha_2}} A_{21} \rho_2^0  \frac{ \left( \lambda_k + p_2\right)^{1-\alpha_2}-p_2^{1-\alpha_2}}{\lambda_k} -f_{21}\right]\left[k^2\frac{\sigma^2_1}{\eta_1^{\alpha_1}} A_{12} \rho_1^0  \frac{\left( \lambda_k + p_1\right)^{1-\alpha_1}-p_1^{1-\alpha_1}}{\lambda_k} -f_{12}\right] .
\EE
 
We note that this expression does not reduce exactly to the one given in \cite{yadav}, see Appendix \ref{yadavcorrection}. The mode with wavenumber $k$ is guaranteed to be unstable if Eq.~(\ref{characteristiceq}) has a real and positive root. 

For $0<\alpha_i <1$, we have $\Delta_k\left(\lambda_k\right) \approx \lambda_k^2$ for large real $\lambda_k$, i.e. $\Delta_k(\lambda_k)$ is positive. Additionally, the function $\Delta_k\left(\lambda_k\right)$ is continuous. Hence, if $\Delta_k\left(0\right) <0$ there must be at least one real positive zero. In other words, $\Delta_k\left(0\right)<0$ is a sufficient condition for instability.  This leads to Eq.~(\ref{stabilitycondition}).
We refer to \cite{yadav} for arguments as to why this is not only a sufficient condition but also necessary.  

\subsection{Correction to the Turing instability criterion in \cite{yadav} and verification in simulations}\label{yadavcorrection}
In this Appendix, we point out a small (but consequential) error which was made in the calculation of the Turing instability in \cite{yadav} and built upon in \cite{yadavmilu}.  We also discuss some of the most important consequences.

In going from Eq.~(29) to (31) in \cite{yadav}, a sign error was made in one of the terms. Eq.~(31) in \cite{yadav} should read 
\begin{align}
u\delta\rho_1\left(k,u\right) &= \delta\rho_1\left(k,t=0\right)+ \frac{\sigma^2_1 k^2 \rho_1^0 p^{1-\alpha} h_\alpha\left(u,p\right) }{u \eta_1^\alpha} \left[ A_1 \delta \rho_1\left(k,u\right) + A_2 \rho_2\left(k,u\right) \right] - \frac{\sigma_1^2 k^2 \left( u+p\right)^{1-\alpha}}{\eta_1^\alpha}\nonumber \\
&+ R^+_{11}\left(\rho^0\right)\delta \rho_1\left(x,t\right)+ R^+_{12}\left(\rho^0\right)\delta \rho_2\left(x,t\right) - R^-_{11}\left(\rho^0\right)\delta \rho_1\left(x,t\right) - R^-_{12}\left(\rho^0\right)\delta \rho_2\left(x,t\right) .\label{correctedequation} 
\end{align}
It is the terms involving $A_1$ and $A_2$ in Eq.~(\ref{correctedequation}) which have the incorrect sign in \cite{yadav}.

In \cite{yadavmilu}, conditions for Turing pattern formation are given in Eqs.~(24) and (31). These are a special case of our more general Eq.~(\ref{fullturingcondition}), in that only one subdiffusing species is considered. Given the  sign error in \cite{yadav}, Eqs. (24) and (31) in \cite{yadavmilu} attract corrections as well.  

Following \cite{yadav}, we denote species 1 as the subdiffusing species with per capita removal rate $p$ and total reaction rate $f$, and species 2 as the normally diffusing species with total reaction rate $g$. We define $\gamma = 1-\alpha$ and $\theta_{\gamma} = \frac{\eta_1^\alpha \sigma_2^2 }{\eta_2 \sigma_1^2 p^{1-\alpha}} = \frac{\hat D_2\left(\boldrho^0\right)}{\hat D_1\left(\boldrho^0\right)}$, the ratio of the effective diffusion coefficients $\hat D$ evaluated in the homogeneous steady state. The condition for Turing instability can be written $\theta_{\gamma}> \theta_{\gamma,c}$, where the critical value $\theta_{\gamma,c}$ can take two forms, depending on whether the activator or the inhibitor is subdiffusing.

For a subdiffusing activator one has
\begin{align}
\theta^{\textrm{act}}_{\gamma,c} =& \frac{1}{f_1^2 p^2} \{ -2f_2 g_1 p^2 +f_1 g_2 p^2 +A_2 f_1 g_1 p \rho_1^0 \gamma - 2 A_1 f_2 g_1 p \rho_1^0 \gamma + A_1 f_1 g_2 p \rho_1^0 \gamma \} \nonumber \\
 &+ 2 \sqrt{g_1 (f_2 g_1 - f_1 g_2)p^2\left(p + A_1 \rho_1^0 \gamma \right)\left[ f_2 \left(p + A_1 \rho_1^0 \gamma \right)- A_2 f_1 \rho_1^0 \gamma\right]}\} , 
\end{align}
where as for the case of a subdiffusing inhibitor the condition reads
\begin{align}
\theta^{\textrm{inh}}_{\gamma,c} =& \Big(\frac{1}{f_1^2 p^2} \{ -2f_2 g_1 p^2 +f_1 g_2 p^2 +A_2 f_1 g_1 p \rho_1^0 \gamma - 2 A_1 f_2 g_1 p \rho_1^0 \gamma + A_1 f_1 g_2 p \rho_1^0 \gamma \} \nonumber \\
 &- 2 \sqrt{g_1 (f_2 g_1 - f_1 g_2)p^2\left(p + A_1 \rho_1^0 \gamma \right)\left[ f_2 \left(p + A_1 \rho_1^0 \gamma \right)- A_2 f_1 \rho_1^0 \gamma\right]}\}\Big)^{-1} . \label{critical}
\end{align}
These expressions are different from the ones in \cite{yadavmilu} in that the signs of $A_1$ and $A_2$ have both been inverted.

\begin{figure*}[t!!!]
 
              \includegraphics[width=0.47\textwidth]{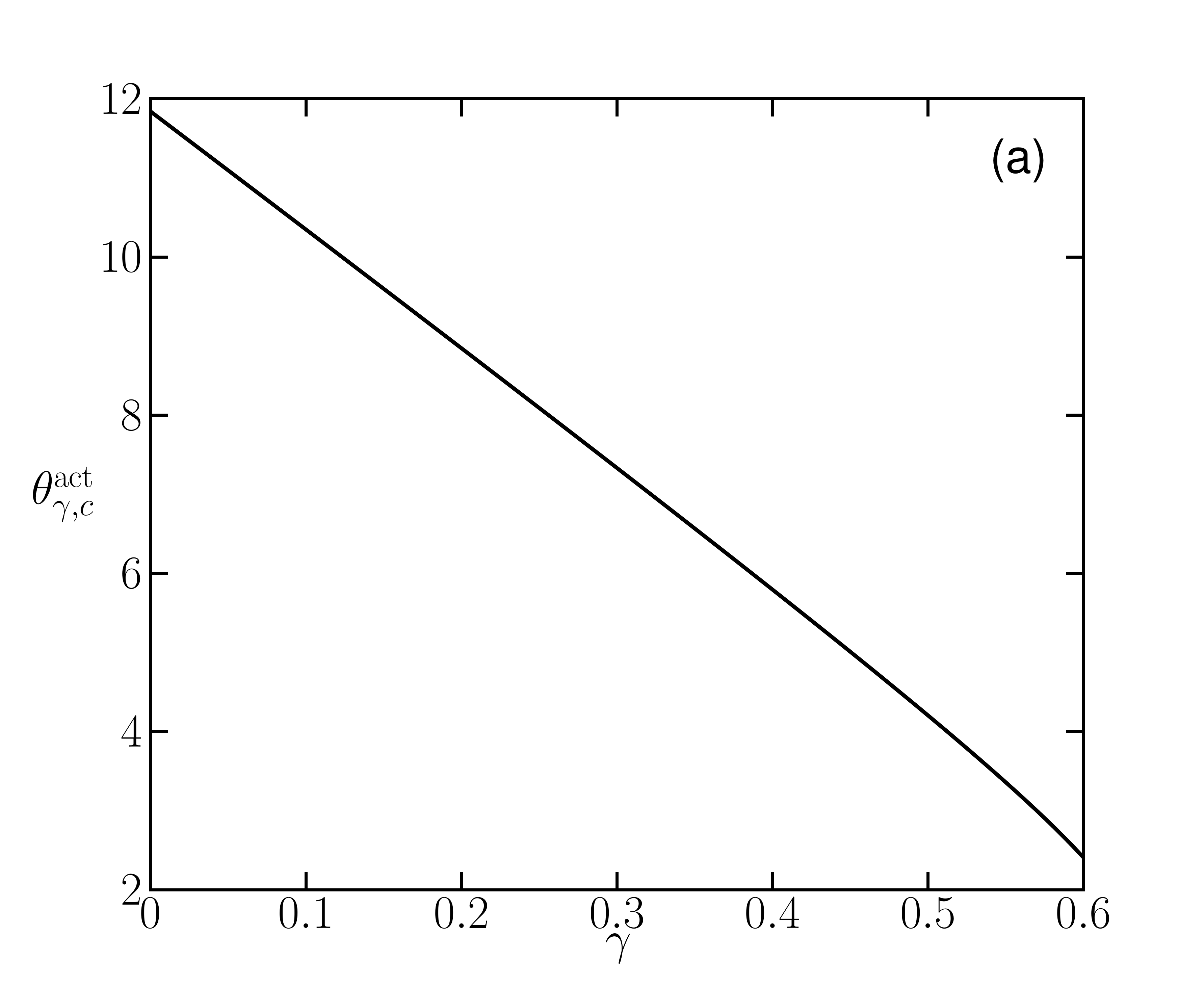}
		\includegraphics[width=0.47\textwidth]{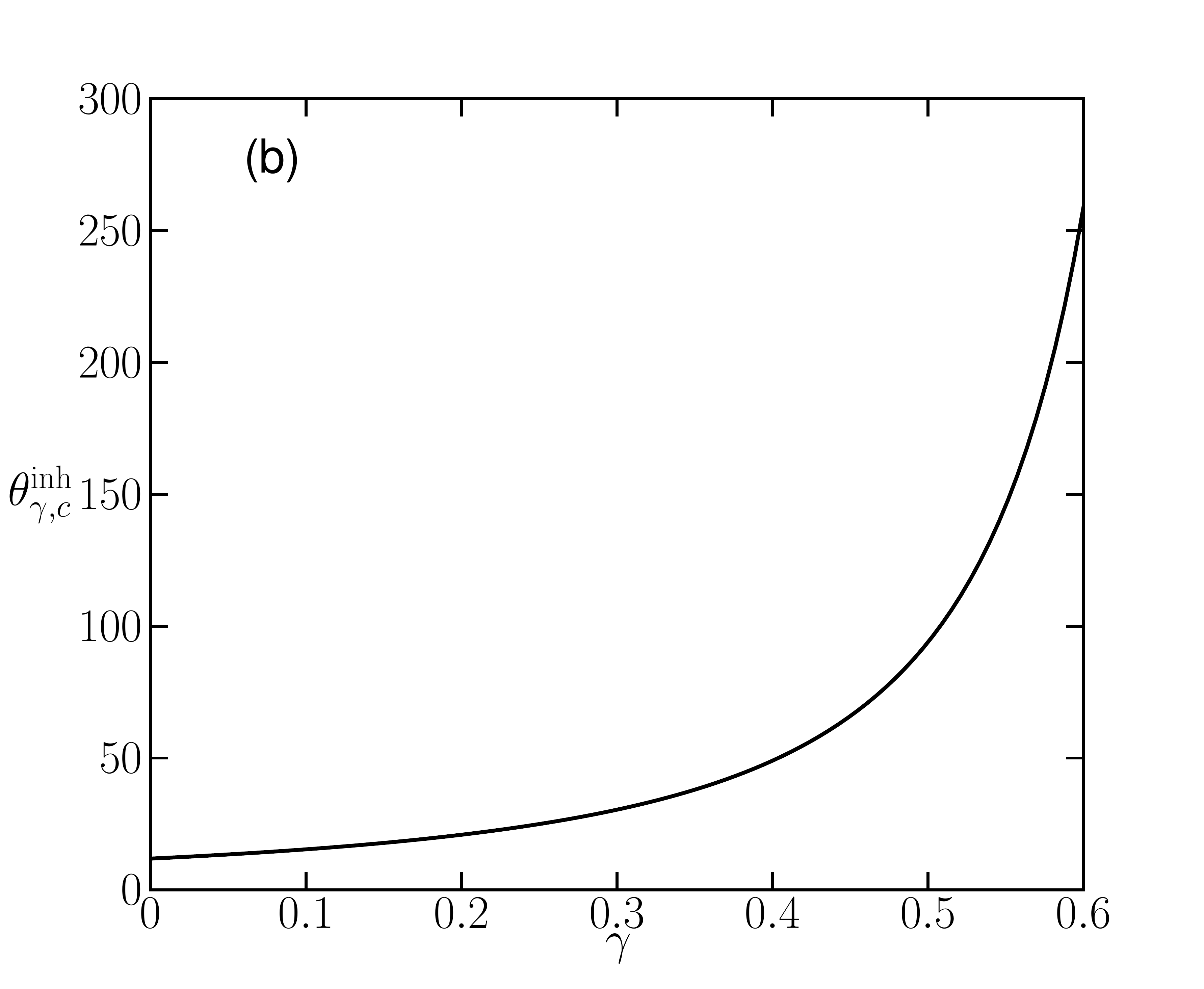}
 
\caption{Value of $\theta_\gamma$ marking the Turing instability in the Lengyel-Epstein model as a function of $\gamma = 1-\alpha$. Model parameters $a' = 50$ and $b' = 40$. (a) and (b) correspond to Figs.~2 and 3 in \cite{yadavmilu} respectively. }
	\label{fig:critical}
\end{figure*}

The sign error has important consequences to the conclusions drawn in \cite{yadavmilu}. In fact, with the error taken into account the phase diagrams shown in Figs. 2 and 3 in \cite{yadavmilu} change considerably. We present the corrected phase diagrams in Fig. \ref{fig:critical}. 
\medskip

{\em Verification in simulations.} To the best of our knowledge the theoretical predictions of \cite{yadavmilu} have not been tested numerically. To verify Eq.~(\ref{critical}) we use the method described in the main text and in Appendix \ref{numericalmethod}. The version of the Lengyel-Epstein model used in \cite{yadavmilu} can be obtained from the setup described in Eqs.~(\ref{le1}) and (\ref{le2}) by the re-scaling $\rho_A' = \frac{\rho_A}{N}, \rho_B' = \frac{c \rho_2}{N}$ and
$d = b = 1 , c = b', a = a'$ (dashed quantities are the ones used in \cite{yadavmilu}).

This yields 
\begin{align}
f'\left(\rho'\right) &= a' - \rho_A' -4 \frac{\rho_A' \rho_B'}{1+\left(\rho_A'\right)^2}, \nonumber \\
g'\left(\rho'\right) &= b' \left( \rho_A' - \frac{\rho_A' \rho_B'}{1+\left(\rho_A'\right)^2}\right), \nonumber \\
\frac{R_A^{'-}}{\rho'_A} &= 1 + 4 \frac{\rho_B'}{1+\left(\rho_A'\right)^2} .
\end{align}

\begin{figure*}[h!!!]
        \includegraphics[width=0.47\textwidth]{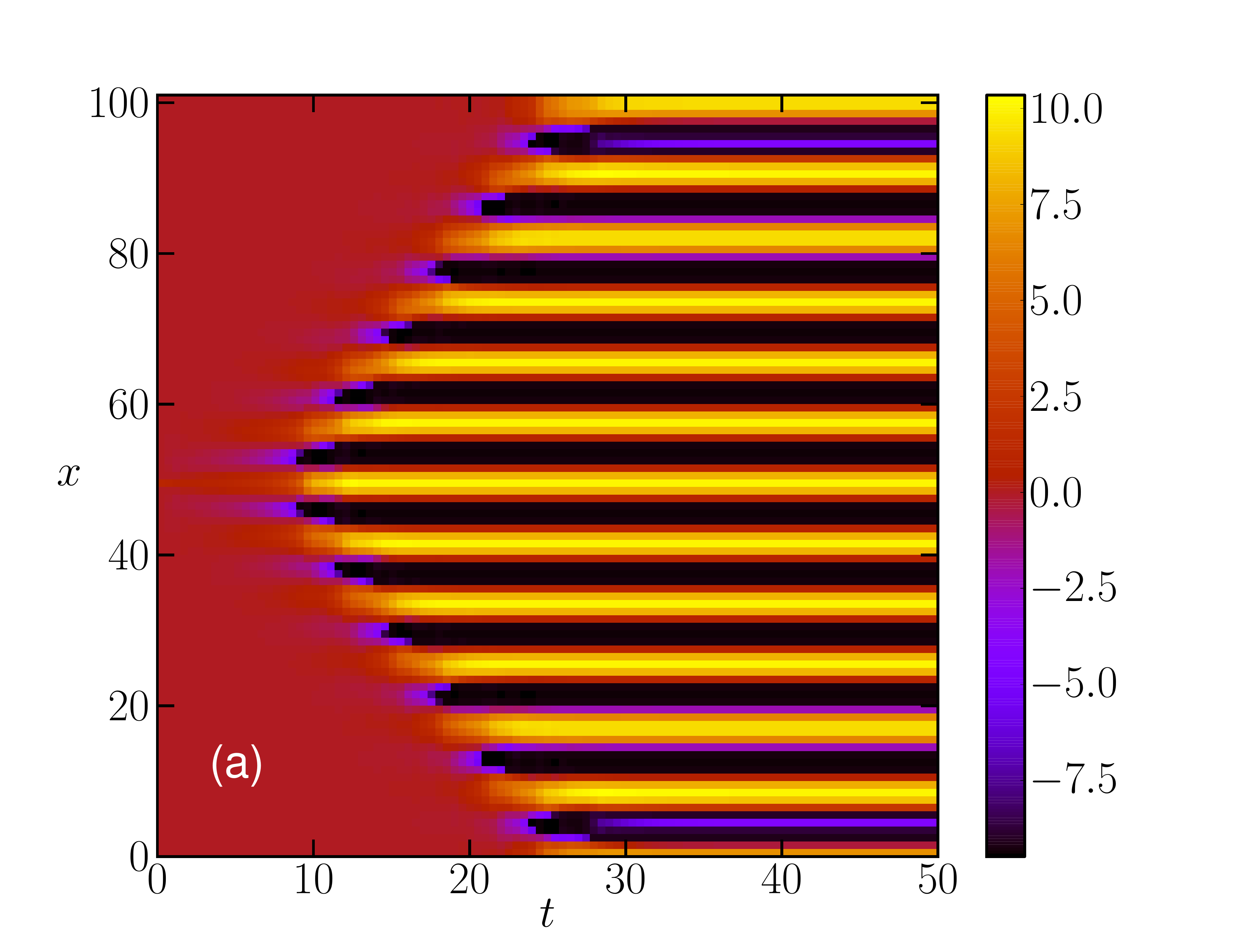}
		\includegraphics[width=0.47\textwidth]{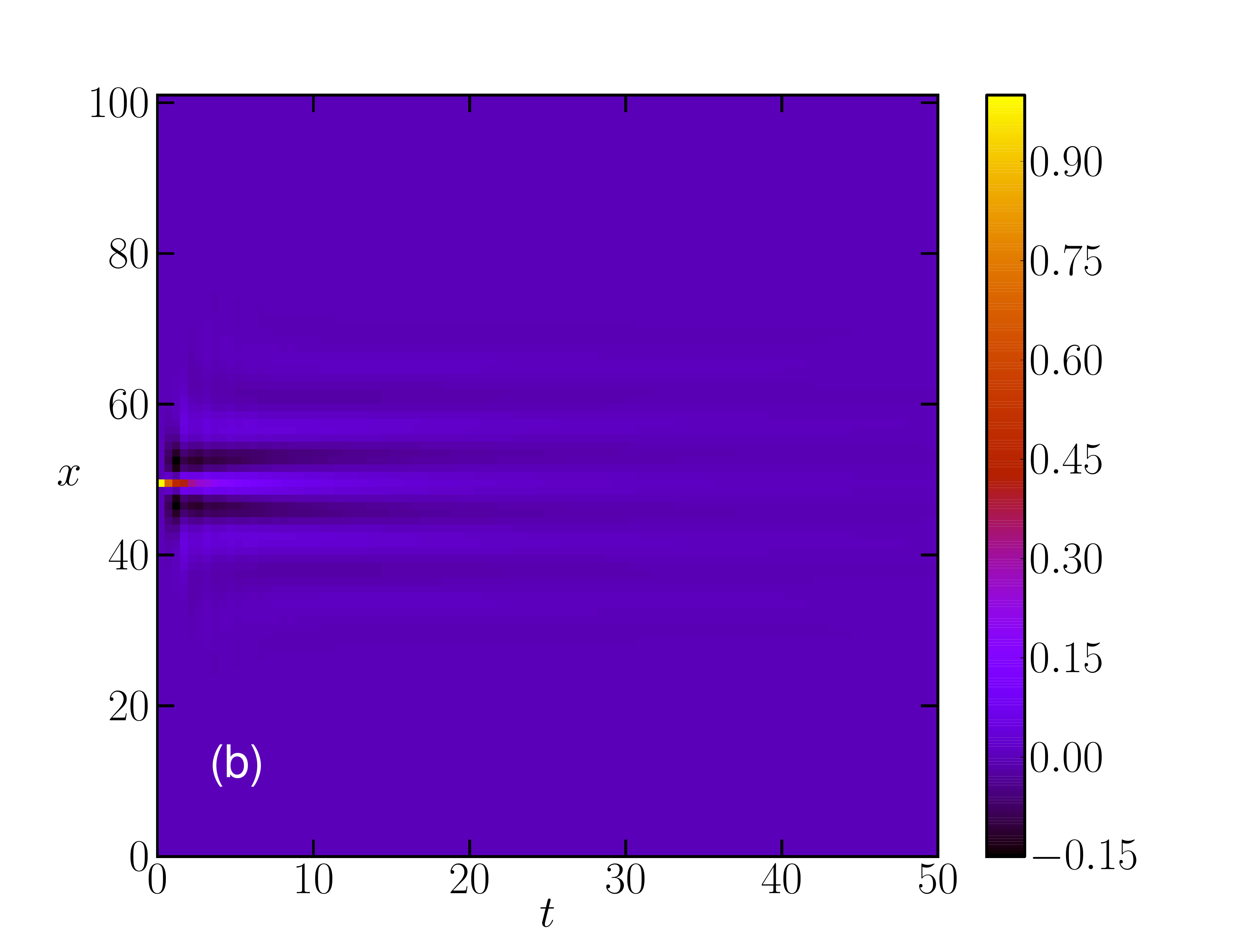}
		\includegraphics[width=0.47\textwidth]{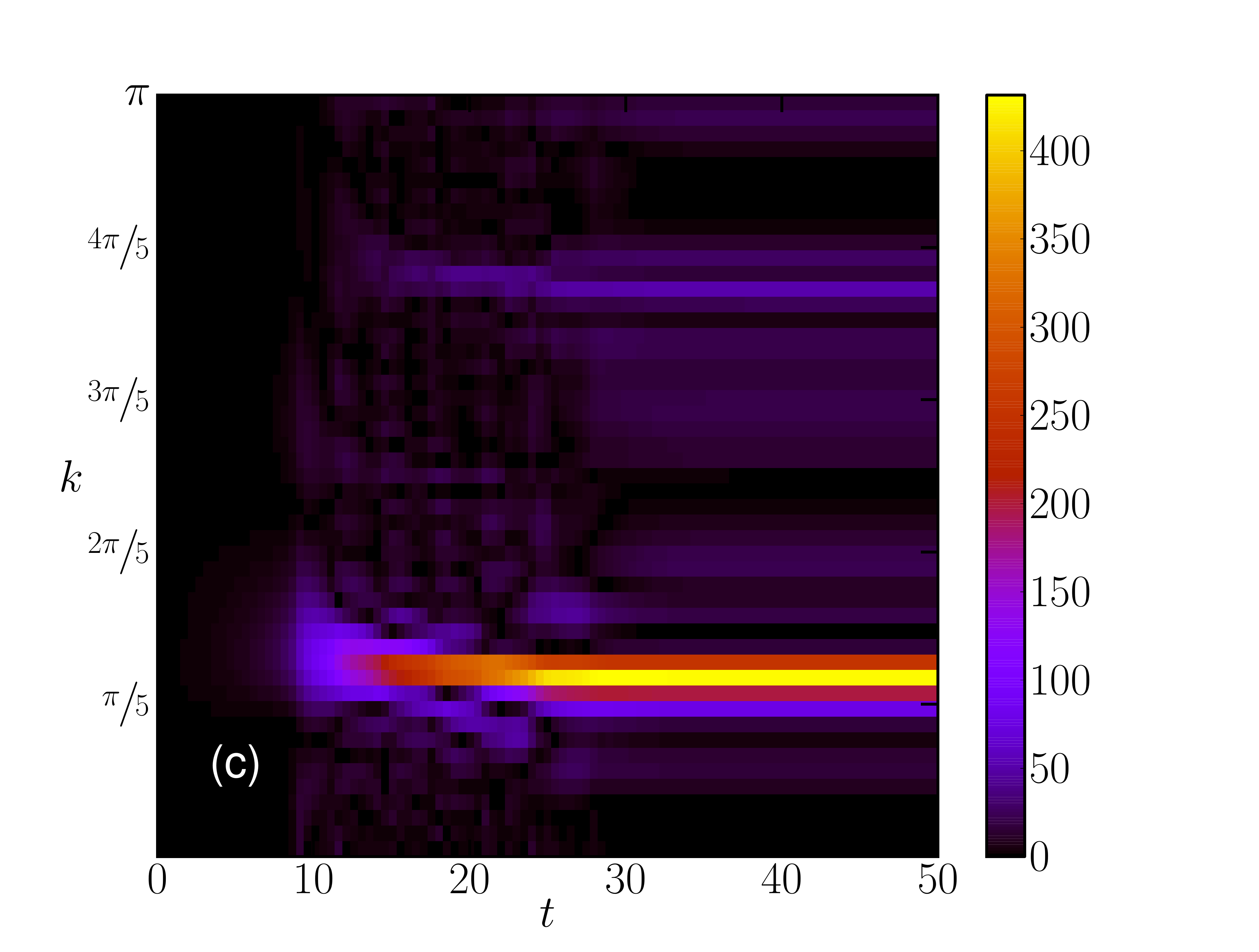}
		\includegraphics[width=0.47\textwidth]{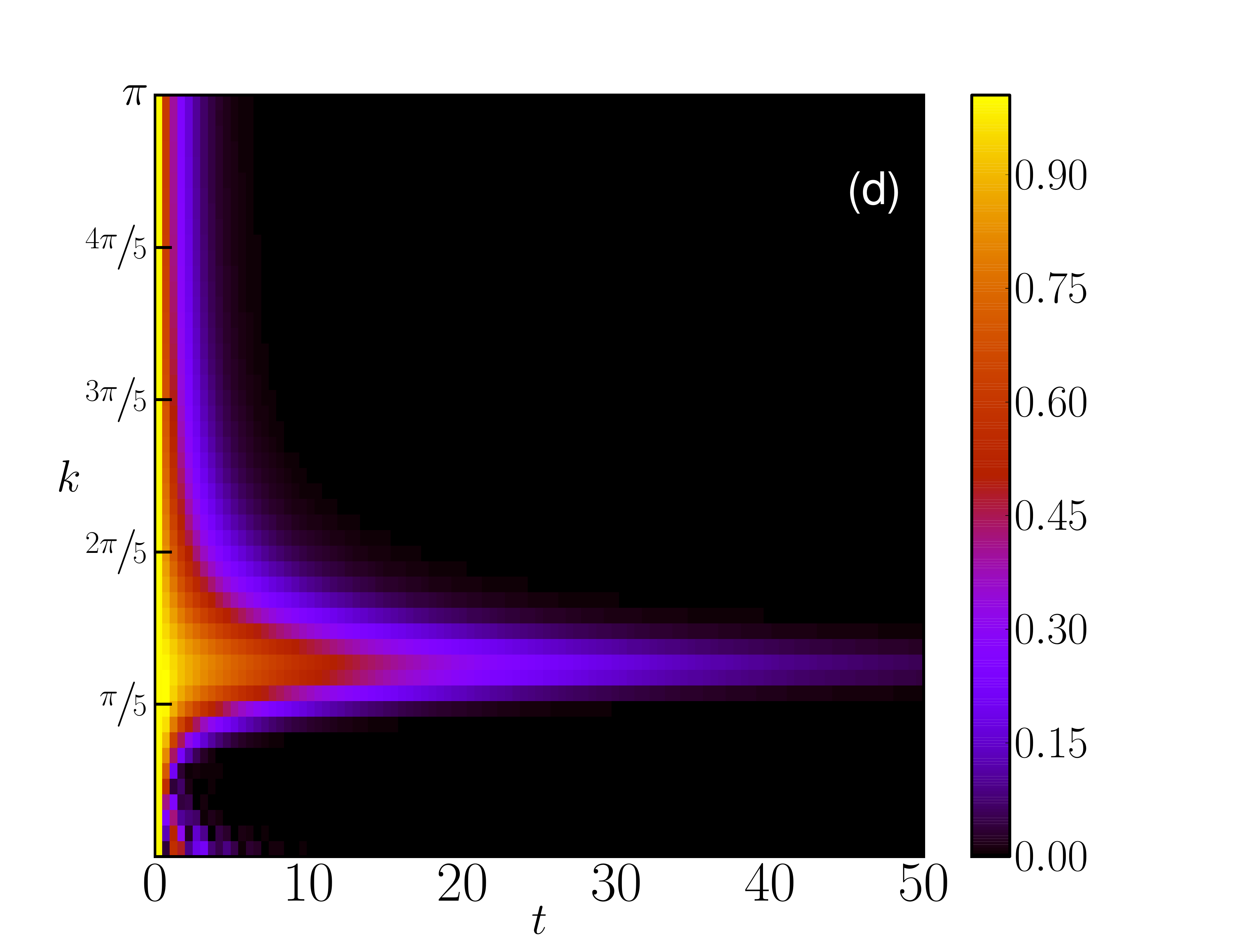}

\caption{Results from numerical integration of the Lengyel-Epstein model with subdiffusing activator, subject to an initial small perturbation around the homogeneous fixed point at $x=51$. The upper panels show deviations from the fixed point in real space, the lower panels are in Fourier space. ((a) and (c)): Model parameters are above the threshold for Turing pattern formation in Fig.~\ref{fig:critical}, but below the threshold predicted in Fig.~2 of \cite{yadavmilu} ($a' = 50, b' = 40, \gamma = 0.5, \eta_1 = 0.02, \eta_2 = 0.01$, resulting in $\theta_\gamma = \eta_1^\alpha/\eta_2 p^{1-\alpha} = 6.32$) . Pattern formation is clearly visible in real space and in Fourier space. ((b) and (d)): Model parameters $a', b'$ and $\gamma$ are as on the left, but $\eta_1 = 0.008, \eta_2 = 0.01$, resulting in $\theta_\gamma = 4$. This is just below the threshold for pattern formation in Fig.~\ref{fig:critical}. The perturbation decays, and no patterns are found.}
	\label{fig:simulations}
\end{figure*}

An example of the outcome of the numerical integration is shown in Fig. \ref{fig:simulations}. One finds patterned stationary states only when $\theta_{\gamma}$ is greater than the critical value given in Eq.~(\ref{critical}). At the same time patterns can be observed below the threshold given in Fig. 2 of \cite{yadavmilu}, confirming our version of the calculation of the phase diagram. 

Importantly, these numerical results demonstrate that, in this model, one can observe Turing patterns for lower values of $\theta_\gamma$ when the activator is subdiffusing than when both reactants are diffusing normally or when the inhibitor subdiffuses. The opposite is reported in \cite{yadavmilu}.

\section{Limit of large removal rates}\label{appendix:fastdeath}
In this Appendix, we derive Eq.~(\ref{approxresult2}). First, we rewrite Eq.~(\ref{fracdiffdef}) in the following form
\begin{align}
{}_0D_t^{1-\alpha} f\left(t\right) &= \frac{1}{\Gamma\left(\alpha\right)} \frac{\partial}{\partial t} \int_0^t \frac{f\left(t-\tau\right)}{\tau^{1-\alpha}}d\tau , \nonumber \\
 &= \frac{1}{\Gamma\left(\alpha\right)} \left[ t^{\alpha -1} f\left(0\right) + \int_0^t\frac{\frac{\partial}{\partial t}f\left(t-\tau\right)}{\tau^{1-\alpha}} d\tau \right] . 
\end{align}
Thus we have
\begin{align}
{}_0D^{1-\alpha_i}_t &\left\{ \rho_i\left(x, t\right) e^{ \int_0^t p_i\left(x,t'\right) dt'}\right\} \nonumber \\
&= \frac{1}{\Gamma\left(\alpha\right)} \Bigg\{ t^{\alpha -1} \rho_i\left(x, 0\right) + \int_0^t\tau^{\alpha-1} \left[ \frac{\partial \rho_i\left(x, t- \tau \right)}{\partial t} +  \rho_i\left(x, t- \tau \right) p_i\left(x, t- \tau \right)\right]e^{ \int_0^{t-\tau} p_i\left(x,t'\right) dt'} d\tau \Bigg\} . \label{expandedfracder}
\end{align}
We now assume that the removal rate for particles of species $i$ at the homogeneous fixed point is large. So, we would like to find a series expansion for integrals of the form
\begin{align}
\int_0^t \tau^{\alpha-1} f\left(t-\tau\right) e^{M g\left(t-\tau\right)} d\tau ,
\end{align}
where $M$ is a large dimensionless parameter. This can be done using a method analogous to Laplace's method \cite{olverasymptotics}.\\
 We presume that the functions $f$ and $g$ can be expanded as Taylor series, which converge for $0<\tau<t$, such that
\begin{align}
f\left(t-\tau\right) = f\left(t\right) - \tau f'\left(t\right) + \frac{1}{2!}\tau^2 f''\left(t\right) + \cdots , \nonumber \\
g\left(t-\tau\right) = g\left(t\right) - \tau g'\left(t\right) + \frac{1}{2!}\tau^2 g''\left(t\right) + \cdots.
\end{align}
In our case, $g'\left(t\right)>0$. Noting that the series for the exponential function is absolutely convergent for any value, one then obtains
\begin{align}
&\int_0^t \tau^{\alpha-1} f\left(t-\tau\right) e^{M g\left(t-\tau\right)} d\tau \nonumber \\
&= \int_0^t \tau^{\alpha - 1} \left[  f\left(t\right) - \tau f'\left(t\right) + \frac{1}{2!}\tau^2 f''\left(t\right) + \cdots \right] \nonumber \\
&\times \left\{1 + M \left[ \frac{1}{2!}\tau^2 g''\left(t\right) - \frac{1}{3!}\tau^3 g'''\left(t\right) + \cdots\right] + \frac{1}{2!}M^2\left[ \frac{1}{2!}\tau^2 g''\left(t\right) - \frac{1}{3!}\tau^3 g'''\left(t\right) + \cdots\right]^2 + \cdots \right\}\nonumber \\
&\times e^{M g\left(t\right)} e^{-Mg'\left(t\right) \tau} d\tau . \label{expanded}
\end{align}
We now note the following 
\begin{align}
\int_0^t \tau^{\alpha + n -1} e^{-M \tau g'\left(t\right)} d\tau = \left[g'\left(t\right) M 
\right]^{- \left(\alpha + n\right)}\gamma\left( \alpha + n, g'\left(t\right) M t\right) , 
\end{align}
where $\gamma(s,x)=\int_0^x t^{s-1} e^{-t} dt$ is the lower incomplete gamma function. Expanding further in Eq.~(\ref{expanded}) and re-organising terms, one obtains the following asymptotic series 
\begin{align}
&\int_0^t \tau^{\alpha-1} f\left(t-\tau\right) e^{M g\left(t-\tau\right)} d\tau \nonumber \\
&= e^{M g\left(t\right)} \left[Mg'\left(t\right)\right]^{-\alpha}\Bigg\{ f\left(t\right)\gamma\left( \alpha, g'\left(t\right) M t\right) + \frac{\left[ \frac{f\left(t \right)g''\left( t\right)}{2!g'\left(t\right)^2}\gamma\left(\alpha + 2, g'\left(t\right)M t \right)- \frac{f'\left(t\right)}{g'\left(t\right)}\gamma\left(\alpha + 1, g'\left(t\right)M t \right)\right]}{M} + \cdots\Bigg\} . \label{asymptoticseries}
\end{align}
The lower incomplete gamma function fulfills following relation \cite{magnusspecialfunctions}
\begin{align}
\gamma\left(s,x\right) =\Gamma\left(s\right) - \Gamma\left(s,x\right) = \Gamma\left(s\right) - e^{-x} x^{s - 1} - e^{-x} x^{s - 1} \sum_{k = 1}^\infty \frac{\prod_{l = 1}^{k} \left(s - l \right)}{x^k} .\label{gammaexpand}
\end{align}
Although the arguments of the lower incomplete gamma functions in Eq.~(\ref{asymptoticseries}) depend on $M$, one can see from Eq.~(\ref{gammaexpand}) that the series expansion for the lower incomplete gamma function consists of a constant term plus a correction term proportional to $e^{-g'\left(t\right)Mt}$. That is, $\gamma\left(\alpha + n,g'\left(t\right)Mt\right) =\Gamma\left(\alpha + n\right) + \mathcal{O}\left(  e^{-g'\left(t\right)Mt} \left[g'\left(t\right)Mt\right]^{\alpha + n - 1}\right)$. This correction always decays more quickly than $M^{\alpha + r}$ for any $r$. Therefore, we can write 
\begin{align}
\int_0^t \tau^{\alpha-1} f\left(t-\tau\right) e^{M g\left(t-\tau\right)} d\tau = e^{M g\left(t\right)} \left[Mg'\left(t\right)\right]^{-\alpha}\Bigg\{ f\left(t\right)\Gamma\left( \alpha\right)  + \mathcal{O}\left(\frac{1}{M} \right) \Bigg\} .
\end{align}
For our problem, we presume that $p_i\left(x,t\right) = M s_i(x,t)$, where $M\gg 1$ represents the limit of large removal rates, and where $s_i(x,t)={\cal O}(M^0)$. We obtain
\begin{align}
{}_0D^{1-\alpha_i}_t &\left\{ \rho_i\left(x, t\right) e^{ \int_0^t p_i\left(x,t'\right) dt'}\right\} \nonumber \\
&= \frac{1}{\Gamma\left(\alpha\right)} \Bigg\{ t^{\alpha -1} \rho_i\left(x, 0\right) + \int_0^t\tau^{\alpha-1} \left[ \frac{\partial \rho_i\left(x, t- \tau \right)}{\partial t} +  \rho_i\left(x, t- \tau \right) p_i\left(x, t- \tau \right)\right]e^{ \int_0^{t-\tau} p_i\left(x,t'\right) dt'} d\tau \Bigg\} , \nonumber \\
&= e^{ \int_0^{t} p_i\left(x,t'\right) dt'} p_i\left(x,t\right)^{1-\alpha} \rho_i\left(x,t\right) \left[ 1 + \mathcal{O}\left(\frac{1}{M}\right)\right] ,
\end{align}
since $\frac{\partial \rho_i\left(x,t\right)}{\partial t} = \mathcal{O}\left(M^0\right)$ and the term $\frac{1}{\Gamma\left(\alpha\right)} t^{\alpha -1} \rho_i\left(x, 0\right)$ is exponentially smaller than the rest. Hence, for large $M$ [or large $p_i\left(x,t\right)$], we arrive at the approximation in Eq.~(\ref{approxresult2}).
\section{Mean squared displacements}\label{appendix:effectiveprocess}
In this Appendix, we derive Eqs.~(\ref{subdiffusivelaw}) and (\ref{diffusivelaw}). We start from Eq.~(\ref{effectiveequation}) and take combined Fourier and Laplace transforms with respect to position and time respectively. We find
\begin{align}
 \hat{\tilde{C}}_i\left(k,u\right) = \frac{1}{u  + \frac{\sigma^2_i}{\eta_i^\alpha} k^2 \left[u +\left( 1-\theta\right) p_i\right]^{1-\alpha_i}} .
\end{align}
Noting that $P_i(x,t)=C_i(x,t)$ for $\theta=0$, one has $\avg{x^2}=\int C_i(x,t) x^2 dx$ for both $\theta=0$ and $\theta=1$. Using $\langle x^2 \rangle = -\frac{\partial^2}{\partial k^2}\tilde{C}_i\left(k,t\right) \vert_{k=0}$, one then obtains
\begin{align}
\mathcal{L}_t\left\{ \langle x^2\left(t\right)\rangle\right\}\left(u\right) = 2 \frac{\sigma^2_i}{\eta_i^\alpha} \frac{\left[u + \left(1-\theta\right) p_i\right]^{1-\alpha_i}}{u ^2} . \label{msdlaplace}
\end{align}
Inverting the Laplace transform one finds for $\theta = 0$ 
\begin{align}
&\langle x^2\left(t\right)\rangle = 2\frac{\sigma^2_i}{\eta_i^\alpha} e^{-p_it}{}_0 D_t^{1-\alpha_i} \left\{ e^{p_it} t \right\} \nonumber \\
&= 2\frac{\sigma^2_i}{\eta_i^\alpha}\frac{1}{\Gamma\left(\alpha_i\right)} p_i^{-\alpha_i}\left[ \left( p_it +1\right) \gamma\left( \alpha_i, p_it \right) - \gamma\left( \alpha_i-1, p_it \right) \right] \nonumber \\
&\approx 2\frac{\sigma^2_i}{\eta_i^\alpha} p_i^{1-\alpha_i} t , 
\end{align}
where we have used an approximation similar to Eq. (\ref{fracderexp}) in the limit $p_it \gg 1$.

 For $\theta = 1$ on the other hand one finds, using Eq.~(\ref{msdlaplace}) and the fact that $\mathcal{L}_t\left(u\right) \left\{ t^\alpha\right\} = \Gamma\left(1+\alpha\right) u^{-\left(1+\alpha\right)}$, 
\begin{align}
\langle x^2\left(t\right)\rangle = 2\frac{\sigma^2_i}{\eta_i^\alpha} \frac{1}{\Gamma\left( 1 + \alpha_i\right)} t^{\alpha_i}.
\end{align}
So one obtains a subdiffusive law for the mean squared displacement when only the surviving particles are included in the ensemble.
\section{Further details on the numerical integration of the reaction-subdiffusion equation}\label{numericalmethod}
The algorithm for the numerical integration of Eq.~(\ref{full}) operates on a lattice with spacing $\Delta x$ and in discretised time with time step $\Delta t$. It can be stated as follows:

\begin{enumerate}
\item[1.] Initialise arrays which can store the histories of particle concentrations for each lattice site as well as the history of the integral $I\left(x,t \right) = \int_0^t \frac{R^-_i \left[\boldrho\left(x,t' \right)\right]}{\rho_i\left(x,t' \right)} dt'$. Choose a starting configuration for the system. Initialise $I\left(x, 0\right) = 0$. Choose the number of steps after which to disregard the contribution to the integration: $j_{\mathrm{cut}}$.\\
\item[2.] For each site $x$, increment the value of the integral $I\left(x,t \right)$ according to 
\be
I\left(x,j \Delta t\right) = I\left[x,\left(j-1\right) \Delta t\right] + \Delta t \frac{R^-_i \left[\rho\left(x,\left(j-1\right) \Delta t\right)\right]}{\rho_i\left[x, \left(j-1\right) \Delta t\right]},
\ee
and keep a record of all previous values of this quantity, up to the cut-off.
\item[3.] For each site $x$, calculate the quantity 
\begin{align}
F\left(x,j \Delta t\right) &= \exp\left(- I\left(x, j \Delta t\right)\right)  {}_0 D_{j \Delta t}^{1-\alpha} \left\{ \exp\left[ I\left(x, j \Delta t\right)\right] \rho_i\left(x,j \Delta t\right)\right\} \nonumber \\
&\approx \frac{1}{\left(\Delta t\right)^\alpha} \sum_{n=\max\left(j - j_{\mathrm{cut}}, 0\right)}^{j} \left( -1\right)^n \binom{\alpha}{n} \exp\left[ -I\left(x,  n\Delta t\right)\right] \rho_i\left[x,\left(j  - n\right)\Delta t\right],
\end{align}
which uses the Gr\"{u}nwald-Letnikov derivative in Eq.~(\ref{glderiv}) and the histories of the quantities $I\left(x,t \right)$, $\rho_i\left(x, t\right)$.\\
\item[4.] Increment the concentrations according to 
\begin{align}
\rho_i\left(x, j \Delta t\right) &= \rho_i\left[x, \left(j-1\right)\Delta t\right] \nonumber \\
&+ \Delta t \frac{\sigma_i^2}{ \eta_i^{\alpha_i}}\left\{  F\left[x + \Delta x, \left(j-1\right) \Delta t\right] + F\left[x - \Delta x, \left(j-1\right) \Delta t\right] -2 F\left[x , \left(j-1\right) \Delta t\right] \right\} \nonumber \\
&+ \Delta t R^+_i\left[\rho\left(x,\left(j-1\right) \Delta t \right)\right] -\Delta t R^-_i \left[\rho\left(x,\left(j-1\right) \Delta t\right)\right] ,
\end{align}
and keep a record of all previous values of these quantities, up to the cut-off.
\item[5.] Go to 2.
\end{enumerate}

With regards to choosing the cut-off time $t_{\mathrm{cut}}$, a simple test as to whether the cut-off is suitably long is to evaluate $e^{-p_i t}{}_0D_t^{1-\alpha_i} \left[ e^{p_i t}\right]$ and ensure that this agrees with the expected analytical result of $p_i^{1-\alpha_i}$ for long times. Using an infinite $t_{\mathrm{cut}}$ would clearly be the most accurate choice of cut-off. That being said, we found that it was always possible to find a finite cut-off time, which greatly increased the efficiency of the calculations and did not interfere with the results. We tested in selected examples that identical results are obtained for all intents and purposes if no cut-off is used.

Additionally, we note that more sophisticated methods of evaluating the Gr\"unwald-Letnikov derivative do exist \cite{macdonald,zeng,li}, but we found that the above method was sufficient for our purposes.
\end{appendix}

\end{document}